\documentclass[12pt,a4paper,onecolumn,notitlepage,british]{scrartcl}
\usepackage[T1]{fontenc}
\usepackage{scrlayer-scrpage}
\usepackage{setspace}
\usepackage{babel}      
\usepackage[centertags, 
            intlimits   
           ]{amsmath}   
\usepackage{amssymb}    
\usepackage{array}      
\usepackage{upgreek}    
\usepackage{cite}       
\usepackage{xspace}     
\usepackage{slantsc}    
\usepackage{makecell}   
\usepackage{multirow}   
\usepackage{bm}         
\usepackage{dsfont}     
\usepackage{accsupp}    
\usepackage{orcidlink}  
\usepackage{units}      
\usepackage{hyperref}   
\hypersetup{
    bookmarksnumbered,      
    colorlinks,             
    allcolors=black,        
    citecolor=[HTML]007f00, 
    urlcolor=[HTML]00007f,  
    linkcolor=[HTML]b70000, 
    pdftitle = {Measurement of jet production in deep inelastic scattering and NNLO determination of the strong coupling at ZEUS},
    pdfauthor = {ZEUS Collaboration},
}

\KOMAoptions{DIV=14,BCOR=0.5cm}
\KOMAoptions{headinclude=false,footinclude=false}
\KOMAoptions{headings=normal,abstract=true}
\KOMAoptions{parskip=half}
\KOMAoptions{numbers=noendperiod}
\renewcommand{\arraystretch}{1.1}
\setkomafont{pageheadfoot}{\normalsize\rmfamily}
\setkomafont{disposition}{\rmfamily\bfseries}
\addtokomafont{section}{\Large}
\addtokomafont{subsection}{\large}
\addtokomafont{subsubsection}{\large}
\setkomafont{caption}{\rmfamily\itshape}
\setkomafont{captionlabel}{\rmfamily\upshape\bfseries}
\setcapindent{0pt}
\deffootnote{1em}{1em}{\textsuperscript{\thefootnotemark}\ }
\KOMAoptions{DIV=last}

\newcommand{\pbi}{\,\text{pb}^{-1}}
\newcommand{\GeV}{{\text{Ge}\kern-0.06667em\text{V\/}}}
\newcommand{\cm}{\,\text{cm}}
\DeclareMathOperator{\diag}{diag}
\newcommand{\abs}[1]{\vert#1\vert}
\newcommand{\unc}[2]{{}^{\raisebox{0.5pt}{\makebox[6pt][c]{\scriptsize$+$}}0.00#1}_{\raisebox{0.5pt}{\makebox[6pt][c]{\scriptsize$-$}}0.00#2}}

\begin{document}

\pdfbookmark[section]{Abstract}{abstract}
\setcounter{page}{0}
\titlehead{\raggedleft\large
  DESY-23-129\par
  September 2023\par
  \vspace*{4.0cm}}
\title{\LARGE Measurement of jet production in deep inelastic scattering and NNLO determination of the strong coupling at ZEUS}
\author{\large ZEUS Collaboration}
\date{\large~\vspace*{2cm}}

\maketitle

\begin{abstract}\noindent
A new measurement of inclusive-jet cross sections in the Breit frame in neutral current deep inelastic scattering using the ZEUS detector at the HERA collider is presented. The data were taken in the years 2004 to 2007 at a centre-of-mass energy of $\unit[318]{\GeV}$ and correspond to an integrated luminosity of $\unit[347]{\pbi}$. The jets were reconstructed using the $k_t$-algorithm in the Breit reference frame. They have been measured as a function of the squared momentum transfer, $Q^2$, and the transverse momentum of the jets in the Breit frame, $p_{\perp,\text{Breit}}$.
The measured jet cross sections are compared to previous measurements and to perturbative QCD predictions. The measurement has been used in a next-to-next-to-leading-order QCD analysis to perform a simultaneous determination of parton distribution functions of the proton and the strong coupling, resulting in a value of $\alpha_s(M_Z^2) = 0.1142 \pm 0.0017 \text{~\small(experimental/fit)}$ $\unc{06}{07} \text{\small~(model/parameterisation)}$ $\unc{06}{04} \text{\small~(scale)}$, whose accuracy is improved compared to similar measurements. In addition, the running of the strong coupling is demonstrated using data obtained at different scales.
\end{abstract}

\thispagestyle{empty}
\clearpage

{\topmargin-1.cm
\evensidemargin-0.3cm
\oddsidemargin-0.3cm
\textwidth 16.cm
\textheight 680pt
\parindent0.cm
\parskip0.3cm plus0.05cm minus0.05cm
\pagenumbering{Roman}
\newcommand{\authorinfo}[3][\textit]{%
\makebox[3ex]{$^{#2}$}
\begin{minipage}[t]{14cm}\raggedright
#1{#3}
\end{minipage}\par
}

\begin{center}
{\Large The ZEUS Collaboration}
\end{center}

{\small\raggedright

I.~Abt$^{1}$, 
R. Aggarwal$^{2}$, 
V.~Aushev$^{3}$, 
O.~Behnke$^{4}$, 
A.~Bertolin$^{5}$, 
I.~Bloch$^{6}$, 
I.~Brock$^{7}$, 
N.H.~Brook$^{8, a}$, 
R.~Brugnera$^{9}$, 
A.~Bruni$^{10}$, 
P.J.~Bussey$^{11}$, 
A.~Caldwell$^{1}$, 
C.D.~Catterall$^{12}$, 
J.~Chwastowski$^{13}$, 
J.~Ciborowski$^{14, b}$, 
R.~Ciesielski$^{4, c}$, 
A.M.~Cooper-Sarkar$^{15}$, 
M.~Corradi$^{10, d}$, 
R.K.~Dementiev$^{16}$, 
S.~Dusini$^{5}$, 
J.~Ferrando$^{4}$, 
B.~Foster$^{15, e}$, 
E.~Gallo$^{17, f}$, 
D.~Gangadharan$^{18, g}$, 
A.~Garfagnini$^{9}$, 
A.~Geiser$^{4}$, 
G.~Grzelak$^{14}$, 
C.~Gwenlan$^{15}$, 
D.~Hochman$^{19}$, 
N.Z.~Jomhari$^{4}$, 
I.~Kadenko$^{3}$, 
U.~Karshon$^{19}$, 
P.~Kaur$^{20}$, 
R.~Klanner$^{17}$, 
U.~Klein$^{4, h}$, 
I.A.~Korzhavina$^{16}$, 
N.~Kovalchuk$^{17}$, 
M.~Kuze$^{21}$, 
B.B.~Levchenko$^{16}$, 
A.~Levy$^{22}$, 
B.~L\"ohr$^{4}$, 
E.~Lohrmann$^{17}$, 
A.~Longhin$^{9}$, 
F.~Lorkowski$^{4}$\orcidlink{0000-0003-2677-3805},
E.~Lunghi$^{23}$,
I.~Makarenko$^{4}$, 
J.~Malka$^{4, i}$, 
S.~Masciocchi$^{24, j}$, 
K.~Nagano$^{25}$, 
J.D.~Nam$^{26}$, 
Yu.~Onishchuk$^{3}$, 
E.~Paul$^{7}$, 
I.~Pidhurskyi$^{27}$, 
A.~Polini$^{10}$, 
M.~Przybycie\'n$^{28}$, 
A.~Quintero$^{25}$, 
M.~Ruspa$^{29}$, 
U.~Schneekloth$^{4}$, 
T.~Sch\"orner-Sadenius$^{4}$, 
I.~Selyuzhenkov$^{24}$, 
M.~Shchedrolosiev$^{4}$, 
L.M.~Shcheglova$^{16}$,
N.~Sherrill$^{30}$,
I.O.~Skillicorn$^{11}$, 
W.~S{\l}omi\'nski$^{31}$, 
A.~Solano$^{32}$, 
L.~Stanco$^{5}$, 
N.~Stefaniuk$^{4}$, 
B.~Surrow$^{26}$, 
K.~Tokushuku$^{25}$, 
O.~Turkot$^{4, i}$, 
T.~Tymieniecka$^{33}$, 
A.~Verbytskyi$^{1}$, 
W.A.T.~Wan Abdullah$^{34}$, 
K.~Wichmann$^{4}$, 
M.~Wing$^{8, k}$, 
S.~Yamada$^{25}$, 
Y.~Yamazaki$^{35}$, 
A.F.~\.Zarnecki$^{14}$, 
O.~Zenaiev$^{4, l}$ 
\newpage

{\setlength{\parskip}{0.4em}
\authorinfo{1}{Max-Planck-Institut f\"ur Physik, M\"unchen, Germany}
\authorinfo{2}{DST-Inspire Faculty, Department of Technology, SPPU, India}
\authorinfo{3}{Department of Nuclear Physics, National Taras Shevchenko University of Kyiv, Kyiv, Ukraine}
\authorinfo{4}{Deutsches Elektronen-Synchrotron DESY, Hamburg, Germany}
\authorinfo{5}{INFN Padova, Padova, Italy~$^{A}$}
\authorinfo{6}{Deutsches Elektronen-Synchrotron DESY, Zeuthen, Germany}
\authorinfo{7}{Physikalisches Institut der Universit\"at Bonn, Bonn, Germany~$^{B}$}
\authorinfo{8}{Physics and Astronomy Department, University College London, London, United Kingdom~$^{C}$}
\authorinfo{9}{Dipartimento di Fisica e Astronomia dell' Universit\`a and INFN, Padova, Italy~$^{A}$}
\authorinfo{10}{INFN Bologna, Bologna, Italy~$^{A}$}
\authorinfo{11}{School of Physics and Astronomy, University of Glasgow, Glasgow, United Kingdom~$^{C}$}
\authorinfo{12}{Department of Physics, York University, Ontario, Canada M3J 1P3~$^{D}$}
\authorinfo{13}{The Henryk Niewodniczanski Institute of Nuclear Physics, Polish Academy of Sciences, Krakow, Poland}
\authorinfo{14}{Faculty of Physics, University of Warsaw, Warsaw, Poland}
\authorinfo{15}{Department of Physics, University of Oxford, Oxford, United Kingdom~$^{C}$}
\authorinfo{16}{Affiliated with an institute covered by a current or former collaboration agreement with DESY}
\authorinfo{17}{Hamburg University, Institute of Experimental Physics, Hamburg, Germany~$^{E}$}
\authorinfo{18}{Physikalisches Institut of the University of Heidelberg, Heidelberg, Germany}
\authorinfo{19}{Department of Particle Physics and Astrophysics, Weizmann Institute, Rehovot, Israel}
\authorinfo{20}{Sant Longowal Institute of Engineering and Technology, Longowal, Punjab, India}
\authorinfo{21}{Department of Physics, Tokyo Institute of Technology, Tokyo, Japan~$^{F}$}
\authorinfo{22}{Raymond and Beverly Sackler Faculty of Exact Sciences, School of Physics, Tel Aviv University, Tel Aviv, Israel~$^{G}$}
\authorinfo{23}{Department of Physics, Indiana University Bloomington, Bloomington, IN 47405, USA}
\authorinfo{24}{GSI Helmholtzzentrum f\"{u}r Schwerionenforschung GmbH, Darmstadt, Germany}
\authorinfo{25}{Institute of Particle and Nuclear Studies, KEK, Tsukuba, Japan~$^{F}$}
\authorinfo{26}{Department of Physics, Temple University, Philadelphia, PA 19122, USA~$^{H}$}
\authorinfo{27}{Institut f\"ur Kernphysik, Goethe Universit\"at, Frankfurt am Main, Germany}
\authorinfo{28}{AGH University of Science and Technology, Faculty of Physics and Applied Computer Science, Krakow, Poland}
\authorinfo{29}{Universit\`a del Piemonte Orientale, Novara, and INFN, Torino, Italy~$^{A}$}
\authorinfo{30}{Department of Physics and Astronomy, University of Sussex, Brighton, BN1 9QH, United Kingdom~$^{I}$}
\authorinfo{31}{Department of Physics, Jagellonian University, Krakow, Poland~$^{J}$}
\authorinfo{32}{Universit\`a di Torino and INFN, Torino, Italy~$^{A}$}
\authorinfo{33}{National Centre for Nuclear Research, Warsaw, Poland}
\authorinfo{34}{National Centre for Particle Physics, Universiti Malaya, 50603 Kuala Lumpur, Malaysia~$^{K}$}
\authorinfo{35}{Department of Physics, Kobe University, Kobe, Japan~$^{F}$}
}
\vspace{3em}

{\setlength{\parskip}{0.4em}
\authorinfo[]{ A}{ supported by the Italian National Institute for Nuclear Physics (INFN)}
\authorinfo[]{ B}{ supported by the German Federal Ministry for Education and Research (BMBF), under contract No.\ 05 H09PDF}
\authorinfo[]{ C}{ supported by the Science and Technology Facilities Council, UK}
\authorinfo[]{ D}{ supported by the Natural Sciences and Engineering Research Council of Canada (NSERC)}
\authorinfo[]{ E}{ supported by the German Federal Ministry for Education and Research (BMBF), under contract No.\ 05h09GUF, and the SFB 676 of the Deutsche Forschungsgemeinschaft (DFG)}
\authorinfo[]{ F}{ supported by the Japanese Ministry of Education, Culture, Sports, Science and Technology (MEXT) and its grants for Scientific Research}
\authorinfo[]{ G}{ supported by the Israel Science Foundation}
\authorinfo[]{ H}{ supported in part by the Office of Nuclear Physics within the U.S.\ DOE Office of Science}
\authorinfo[]{ I}{supported in part by the Science and Technology Facilities Council grant number ST/T006048/1}
\authorinfo[]{ J}{supported by the Polish National Science Centre (NCN) grant no.\ DEC-2014/13/B/ST2/02486}
\authorinfo[]{ K}{ supported by HIR grant UM.C/625/1/HIR/149 and UMRG grants RU006-2013, RP012A-13AFR and RP012B-13AFR from Universiti Malaya, and ERGS grant ER004-2012A from the Ministry of Education, Malaysia}
}
\pagebreak[4]

{\setlength{\parskip}{0.4em}
\authorinfo[]{ a}{now at University of Bath, United Kingdom}
\authorinfo[]{ b}{also at Lodz University, Poland}
\authorinfo[]{ c}{now at Rockefeller University, New York, NY 10065, USA}
\authorinfo[]{ d}{now at INFN Roma, Italy}
\authorinfo[]{ e}{also at DESY and University of Hamburg, Hamburg, Germany and supported by a Leverhulme Trust Emeritus Fellowship}
\authorinfo[]{ f}{also at DESY, Hamburg, Germany}
\authorinfo[]{ g}{now at University of Houston, Houston, TX 77004, USA}
\authorinfo[]{ h}{now at University of Liverpool, United Kingdom}
\authorinfo[]{ i}{now at European X-ray Free-Electron Laser facility GmbH, Hamburg, Germany}
\authorinfo[]{ j}{also at Physikalisches Institut of the University of Heidelberg, Heidelberg,  Germany}
\authorinfo[]{ k}{also supported by DESY, Hamburg, Germany}
\authorinfo[]{ l}{now at Hamburg University, II. Institute for Theoretical Physics, Hamburg, Germany }
}
}
}
\clearpage

\pagenumbering{arabic}
\section{Introduction}
\label{sec-int}
The measurement of jet production in $e^\pm p$ scattering is important for the understanding of quantum chromodynamics (QCD) and is a well-established tool to test perturbative QCD predictions\cite{%
pl:b547:164,
pl:b507:70,
epj:c19:289,
pl:b515:17,
epj:c44:183,
hep-ex/0608048,
h1highq2oldjets,
zeusdijets,
h1highq2newjets
}.
Studies of inclusive-jet production, in which each jet is considered individually, in neutral current (NC) deep inelastic scattering (DIS) events are especially suited for precision determinations of the strong coupling, $\alpha_s$, together with the gluon distribution function of the proton. Compared to dijet measurements, inclusive-jet measurements have a smaller statistical uncertainty and smaller associated theoretical uncertainties, as expected for a more inclusive process. Inclusive-jet measurements also give access to a more unconstrained kinematic region, since they are unaffected by the infrared sensitivity that excludes some regions of dijet measurements\cite{np:b507:315}.

\begin{figure}[p]
    \pdfbookmark[section]{Figures}{fig}
    \vfill
    \begin{center}
        \includegraphics{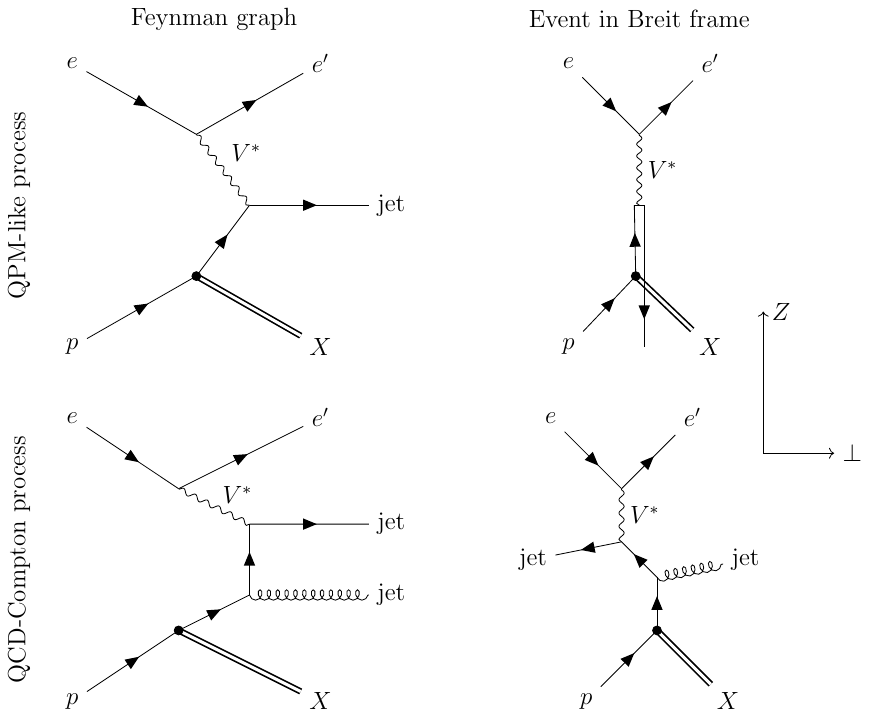}
    \end{center}
    \caption{Single-jet production via the QPM-like process (top row) and dijet production via the QCD-Compton process (bottom row). The left column depicts the Feynman graphs corresponding to each interaction with time running from left to right. The right column depicts the same graphs, arranged in such a way that the directions of the particle lines correspond to the direction of the particle momenta in the longitudinal and radial directions in the Breit frame of reference. The labels $e$, $e'$, $p$, $X$ and $V^*$ denote the incoming and scattered electron, the incoming proton, the proton remnant and the exchanged boson, respectively.}
    \label{fig:breit frame}
    \vfill
\end{figure}

For the study of QCD processes in DIS, the Breit frame of reference has several advantages\nobreak\cite{zeusdijets}.
In this frame, the exchanged virtual boson $V^*$ (a photon or $Z$ boson) collides collinearly with an incoming parton in the proton.
The single-jet production process of the type $V^*q \rightarrow q$, referred to as the quark-parton-model-like (QPM-like) process, is predominantly of zeroth order in $\alpha_s$ and is therefore not of interest for the present analysis.
When viewing this process in the Breit frame, the outgoing quark is scattered back along the collision axis and can therefore be suppressed by selecting jets with a high transverse momentum, $p_{\perp,\text{Breit}}$, with respect to this axis.
This suppression is beneficial for the determination of $\alpha_s$.
In dijet or multi-jet production processes, which do involve hard QCD interactions of order $\alpha_s$ or higher, jets have in general a non-zero transverse momentum in the Breit frame.
The leading-order contributions in the Breit frame are from the QCD-Compton ($V^*q \rightarrow gq$) and boson-gluon-fusion ($V^*g \rightarrow q\bar{q}$) processes.
The Feynman graphs and corresponding depictions of single-jet and multi-jet events in the Breit frame are illustrated in Fig.~\ref{fig:breit frame}.

In this paper, a measurement of double-differential inclusive-jet cross sections in $Q^2$ and $p_{\perp,\text{Breit}}$ in NC DIS events using the ZEUS detector at HERA is presented, where $Q^2$ is the negative square of the four-momentum of the virtual exchanged boson and $p_{\perp,\text{Breit}}$ is the transverse momentum of each jet in the Breit reference frame. The analysis was performed for the kinematic region $Q^2 > \unit[150]{\GeV^2}$ and $p_{\perp,\text{Breit}} > \unit[7]{\GeV}$.

The measured cross sections were used in a QCD analysis at next-to-leading-order (NLO, $\mathcal{O}(\alpha_s^2)$) and next-to-next-to-leading-order (NNLO, $\mathcal{O}(\alpha_s^3)$) to determine the strong coupling, $\alpha_s$. Different strategies for evaluating the scale uncertainty on the measured value are discussed and compared.
Being subject to renormalisation, $\alpha_s$ depends on the scale at which it is evaluated\cite{altarelli2013qcd}. In addition to the global determination of $\alpha_s(M_Z^2)$, a second analysis was performed to investigate its running by determining $\alpha_s(\mu^2)$ at different scales, $\mu$.

\section{Experimental set-up}
\label{sec-exp}
The data used in this analysis were collected in the years 2004--2007 and correspond to an integrated luminosity of $\unit[190]{\pbi}$ and $\unit[157]{\pbi}$ for $e^-p$ and $e^+p$ data, respectively\footnote{From here on, in this paper, the term `electron' refers to both electrons and positrons, unless otherwise stated.}. During this period, HERA operated with a proton beam energy of $E_p=\unit[920]{\GeV}$ and an electron beam energy of $E_e=\unit[27.5]{\GeV}$. This corresponds to a centre-of-mass energy of $\sqrt{s}=\unit[318]{\GeV}$.

A detailed description of the ZEUS detector can be found elsewhere~\cite{zeus:1993:bluebook}. A brief outline of the components that are most relevant for this analysis is given below.

In the kinematic range of the analysis, charged particles were tracked in the central tracking detector (CTD)~\cite{nim:a279:290,npps:b32:181,nim:a338:254} and the microvertex detector (MVD)~\cite{nim:a581:656}. These components operated in a magnetic field of \unit[1.43]{\,\text{T}} provided by a thin superconducting solenoid. The CTD consisted of 72~cylindrical drift-chamber layers, organised in nine superlayers covering the polar-angle\footnote{The ZEUS coordinate system is a right-handed Cartesian system, with the $Z$ axis pointing in the nominal proton beam direction, referred to as the ``forward direction'', and the $X$ axis pointing towards the centre of HERA. The coordinate origin is at the centre of the CTD. The pseudorapidity is defined as $\eta_\text{lab}=-\ln\left(\tan\frac{\theta}{2}\right)$, where the polar angle, $\theta$, is measured with respect to the $Z$ axis. The azimuthal angle, $\phi$, is measured with respect to the $X$ axis.} region \mbox{$15^\circ < \theta< 164^\circ$}.
The MVD silicon tracker consisted of a barrel (BMVD) and a forward (FMVD) section. The BMVD contained three layers and provided polar-angle coverage for tracks from $30^\circ$ to $150^\circ$. The four-layer FMVD extended the polar-angle coverage in the forward region to $7^\circ$. After alignment, the single-hit resolution of the MVD was $\unit[24]{\upmu\text{m}}$. The transverse distance of closest approach (DCA) of tracks to the nominal vertex in $X$--$Y$ was measured to have a resolution, averaged over the azimuthal angle, of $\unit[(46 \oplus 122 / p_{T})]{\upmu\text{m}}$, with $p_{T}$ in $\unit{\GeV}$. For CTD-MVD tracks that pass through all nine CTD superlayers, the momentum resolution was $\sigma(p_{T})/p_{T} = 0.0029 p_{T} \oplus 0.0081 \oplus 0.0012/p_{T}$, with $p_{T}$ in $\unit{\GeV}$.

The high-resolution uranium--scintillator calorimeter (CAL)~\cite{nim:a309:77,nim:a309:101,nim:a321:356,nim:a336:23} consisted of three parts: the forward (FCAL), the barrel (BCAL) and the rear (RCAL) calorimeters. Each part was subdivided transversely into towers and longitudinally into one electromagnetic section (EMC) and either one (in RCAL) or two (in BCAL and FCAL) hadronic sections (HAC). The smallest subdivision of the calorimeter was called a cell. The CAL energy resolutions, as measured under test-beam conditions, were $\sigma(E)/E=0.18/\sqrt{E}$ for electrons and $\sigma(E)/E=0.35/\sqrt{E}$ for hadrons, with $E$ in $\unit{\GeV}$.

The position of electrons scattered at small angles to the electron-beam directions was determined with the help of RHES\cite{nim:a277:176}, which consisted of a layer of approximately $10\,000$ ($\unit[2.96\times 3.32]{\cm^2}$) silicon-pad detectors inserted in the RCAL at a depth of 3.3 radiation lengths.

The luminosity was measured using the Bethe-Heitler reaction $ep\,\rightarrow\, e\gamma p$ by a luminosity detector which consisted of independent lead-scintillator calorimeter\cite{desy-92-066,zfp:c63:391,acpp:b32:2025} and magnetic spectrometer\cite{nim:a565:572} systems. The fractional systematic uncertainty on the measured luminosity was $\unit[1.9]{\%}$.

\section{Monte Carlo simulations}
The response of the detector to DIS events with hadron jets was modelled using Monte Carlo (MC) samples. These samples were used to determine the efficiency of the event selection, to estimate the amount of migration of events and jets across bin boundaries, to calibrate the electron- and jet-energy scales and to estimate background contributions. The equivalent luminosity of the signal MC samples exceeded that of the data by at least a factor of seven over the entire kinematic region.

The NC DIS events were generated using matrix elements calculated with the \textsc{Heracles~4.5} program\cite{cpc:69:155} and \textsc{CTEQ5D} parton distribution functions (PDFs)\cite{epj:c12:375}. The calculation included radiative corrections (single photon emission from initial- or final-state lepton, self-energy corrections to the exchanged boson, vertex corrections of the lepton-boson vertex). The simulation of events was done at leading order + parton showering in QCD (LO+PS). Two samples were generated using different models for parton showering. For this purpose, either \textsc{Ariadne~4.12}\cite{cpc:71:15} or \textsc{Lepto~6.5}\cite{cpc:101:108} was used. These two programs implement different variants of a leading-log parton shower. In both samples, hadronisation of the final-state partons was modelled using the Lund string model\cite{prep:97:31} as implemented in \textsc{Jetset 7.410}\cite{cpc:82:74}. These two MC samples are referred to as \textsc{Ariadne} and \textsc{Lepto}, respectively.

The response of the ZEUS detector to the generated events was simulated using the \textsc{Geant~3.21} program\cite{geant}. The simulated events were subjected to the same trigger configurations as the data and were processed using the same reconstruction and analysis algorithms. Physical quantities from events prior to being passed through the detector simulation are referred to as hadron-level quantities. The quantities determined after detector simulation are referred to as detector-level quantities.

The \textsc{Ariadne} and \textsc{Lepto} signal samples were generated in the region $Q^2 > \unit[100]{\GeV^2}$. In addition, a low-$Q^2$ ($\unit[4]{\GeV^2} < Q^2 < \unit[100]{\GeV^2}$) \textsc{Ariadne} sample was generated to estimate the contribution of events that migrate into the signal region.
The background from photoproduction ($Q^2 < \unit[4]{\GeV^2}$) was estimated using a MC sample generated with the \textsc{Herwig~5.9} program\cite{hep-ph-9912396}.

\section{Event selection and reconstruction}
\label{sec:selection}
\subsection{Online selection}
Online event selection was performed using a three-level trigger system\cite{zeus:1993:bluebook,Smith:1992im}.
At the first level, only coarse calorimeter and tracking information was available. Events were selected if they had an energy deposit in the CAL consistent with an isolated electron. Events were also selected if they deposited a large amount of energy in the electromagnetic part of the calorimeter in coincidence with a CTD/MVD track.
At the second level, a requirement on the difference between the total energy and the total longitudinal momentum of the event was used to select NC DIS events. Timing information from the CAL was used to reject events inconsistent with the bunch-crossing time.
At the third level, NC DIS events were accepted based on the identification of a scattered-electron candidate using localised energy deposits in the CAL. These requirements were similar to, but looser than, the offline selection described below.

\subsection{Offline selection of inclusive DIS events}
Candidates for the scattered DIS electron were identified offline using an algorithm that combined information from the CAL, the RHES and the CTD\cite{epj:c11:427}, and the most probable candidate was selected. The kinematic quantities $Q^2$ and the inelasticity, $y$, were reconstructed with the double-angle method\cite{proc:hera:1991:23,HERAPDF20NLO}, also using the hadronic system\cite{thesis:lorkowski}. They are denoted as $Q^2_\text{DA}$ and $y_\text{DA}$.

The reconstructed kinematic region selected for this analysis is $Q^2_\text{DA} > \unit[150]{\GeV^2}$ and $0.2 < y_\text{DA} < 0.7$. The lower limit on the inelasticity removed a region in which hadronisation effects of the jets become large and cannot be simulated reliably. The upper limit ensured a good detector acceptance. Events were selected if they satisfied the following quality criteria:
\begin{itemize}
    \item the presence of a scattered electron candidate was required. This candidate was required to have an energy of $E_e'>\unit[10]{\GeV}$, which ensured a high efficiency of the electron finder and increased the purity of the DIS sample by suppressing background from photoproduction events. The sum of all energy deposits within a cone of radius $0.8$ in the $(\eta_\text{lab}-\phi)$-plane, centred on the electron candidate, was computed, including the energy of the electron candidate itself. The event was rejected if more than $10\%$ of this energy was not assigned to the electron candidate. This requirement removed events in which a jet closely overlaps with the electron;
    \item the difference between the energy, $E$, and the longitudinal momentum, $p_Z$, summed over all detected energy deposits\cite{epj:c11:427} was required to fulfil $\unit[38]{\GeV} < \sum(E-p_Z) < \unit[65]{\GeV}$. This quantity is especially effective in rejecting events in which particles escaped into the rear beam pipe, such as the scattered electron in a photoproduction event or a hard bremsstrahlung photon radiated from the initial-state electron;
    \item the electron candidate was required to have a track associated with it. The momentum of this track had to fulfil $p_\text{track} > \unit[3]{\GeV}$ and it had to intersect the CAL surface no further than $\unit[10]{\cm}$ from the electron candidate. This requirement rejected events in which a photon was misidentified as an electron;
    \item at least one track associated with the primary vertex was required. This track had to have a transverse momentum of at least $\unit[0.2]{\GeV}$ and had to pass through at least three superlayers of the CTD. The fit of the primary vertex had to have a $\chi^2$ per degree of freedom of no more than $10$. These requirements ensured that the position of the primary vertex was well measured;
    \item the longitudinal position of the primary vertex, $Z_\text{vertex}$, was required to be within $\unit[30]{\cm}$ of the nominal $ep$ interaction point. This condition suppressed background events from beam-gas interactions and ensured a high reconstruction efficiency;
    \item the total transverse momentum of the event was required to be consistent with zero by demanding $\abs{\vec{p}_T}/\sqrt{E_T} < \unit[2.5]{\sqrt{\GeV}}$, where $\vec{p}_T$ is the vectorial sum and $E_T$ the scalar sum of all energy deposits in the CAL. This requirement removed background from cosmic-ray interactions and charged current events;
    \item events were rejected if a second isolated energy deposit in the EMC\cite{epj:c11:427} was present, which fulfilled the following two criteria with respect to the DIS electron: azimuthal separation $\Delta\phi > 3$ and energy within $20\%$. However, an event was only rejected if it had less than $\unit[3]{\GeV}$ additional energy deposited in the CAL. This rejected elastic QED-Compton scattering events\cite{thesis:januschek};
    \item events were rejected if the DIS electron was found in certain regions of the detector where the reconstruction of electrons was poor. These regions are the gaps between the CAL components with $Z$ just below $\unit[-98.5]{\cm}$ or just above $\unit[164]{\cm}$, a support pipe in the RCAL $\vert X \vert < \unit[12]{\cm}$ and $Y > \unit[80]{\cm}$ and the outer region of the RCAL with radius $R_\text{RCAL} > \unit[175]{\cm}$.
\end{itemize}

\subsection{Jet reconstruction and selection}
\label{subsec:jet selection}
The concept of the present analysis requires jets to be defined in the Breit frame, see Fig.~\ref{fig:breit frame}.
Constructing the Breit frame requires knowledge of the four-momentum of the exchanged boson. This was computed from the four-momentum of the scattered electron, as obtained from the double-angle method\cite{proc:hera:1991:23,HERAPDF20NLO}. This method does not assume massless partons or jets\cite{thesis:lorkowski}.

Detector-level jets were reconstructed in the Breit frame using the $k_t$-algorithm with the radius parameter set to $R=1$\cite{pr:d48:3160,Cacciari_2012}. For high-energy jets at HERA, it was established that the $k_t$- and anti-$k_t$-algorithms have a very similar performance\cite{CmpJetAlg}. To be consistent with the available theoretical predictions, the $p_t$-weighted recombination scheme was used to obtain massless jets\cite{np:b406:187}. The input to the algorithm was a list of all energy deposits in the CAL above the noise threshold, excluding those associated with the scattered DIS electron and those directly adjacent to the beam pipe. This ensured a uniform response, resolution and calibration throughout the detector. Each energy deposit was treated as equivalent to that of a massless particle. The four-momentum of each energy deposit was boosted to the Breit frame, where the jet reconstruction was performed. The four-momenta of the jets were then also boosted back into the laboratory frame for further correction and selection.

A variety of selection criteria were applied in the Breit and laboratory frames.
An event was rejected if either of the following conditions applied:
\begin{itemize}
    \item a jet was found with a transverse momentum in the Breit frame of $p_{\perp,\text{Breit}} > 5\,\text{GeV}$ and a distance to the electron of less than one unit in the $(\eta_\text{lab}-\phi)$-plane in the laboratory system. This cut further rejected events in which the DIS electron overlapped with a jet;
    \item a jet with $p_{\perp,\text{Breit}} > 5\,\text{GeV}$ was found at $\eta_\text{lab} < -1.5$. This requirement removed events in which a bremsstrahlung photon from the initial-state electron was identified as a jet because such a photon influenced the reconstruction of the kinematic quantities of the event and thereby distorted the construction of the Breit frame\cite{thesis:moritz:2001}.
\end{itemize}

The following requirements were then made for a jet to be accepted in a given event:
\begin{itemize}
    \item a jet whose transverse momentum in the laboratory frame, $p_{T,\text{lab}}$, was less than $\unit[3]{\GeV}$ was rejected owing to the large uncertainty on the energy measurement in the CAL for such jets;
    \item the jet was required to satisfy $-1 < \eta_\text{lab} < 2.5$. The upper cut rejected jets in the very forward direction because there was a high probability that parts of these jets escaped down the beam pipe. The lower cut restricted the measurement to a region of sufficient statistics;
    \item the jet was required to satisfy $p_{\perp,\text{Breit}} > \unit[4.5]{\GeV}$. This requirement excluded jets originating from QPM-like interactions. This range was wider than that used in the measurement so that under- and overflow bins could be included in the unfolding.
\end{itemize}

\section{Corrections to data and simulation}
\label{sec:corrections}
A series of corrections were applied to improve the resolution of reconstructed quantities and to ensure that the MC samples were suitable to unfold the data\cite{thesis:lorkowski}. Corrections to the MC hadron-level distributions (defined in Section~\ref{sec:unfolding}) were derived by comparing the detector-level distributions of data and MC, separately for the \textsc{Ariadne} and \textsc{Lepto} samples. For detector effects, common correction factors were derived by comparing the average of both MC samples to data. The selection described in the previous section was applied to the corrected samples.

\subsection{Corrections to inclusive DIS events}
The following corrections were applied:
\begin{itemize}
    \item the data were recorded with a polarised electron beam with an average polarisation of $0.01$. Weights were applied such that the data correspond to an unpolarised sample. These correction factors were derived as a function of $Q^2_\text{DA}$ using dedicated MC samples;
    \item the efficiency of reconstructing tracks and associating them to the electron candidates was not perfectly described in the MC\cite{thesis:januschek}. Correction factors were derived in the laboratory frame as a function of the azimuthal angle of the scattered DIS electron and applied to the MC samples;
    \item a veto on the fraction of tracks not associated with the vertex was applied at trigger level. The efficiency of this tracking veto was not perfectly described by the MC\cite{thesis:behr}. Correction factors were derived as a function of the trigger-level track multiplicity and applied to the MC samples;
    \item as the energy of the scattered DIS electron was used to construct the Breit frame, its measurement needed to be accurately calibrated at detector level. A MC study was performed to derive the average difference between the measured and true energy of the electron as a function of the azimuthal angles of the electron and the hadronic system. The correction was applied to the measured energy in data and MC.
\end{itemize}

\subsection{Corrections to jets}
\label{subsec:corr:jets}
The following corrections were applied to the inclusive-jet samples:
\begin{itemize}
    \item the measured energy of hadrons was not necessarily perfectly described by the detector simulation\cite{thesis:behr}. Correction factors and offsets were applied to the reconstructed jet energies in the MC as a function of the pseudorapidity of each jet in the laboratory frame and were propagated into the Breit frame. These factors were derived by studying events with only one `hard' jet and comparing the ratios of the transverse momentum of the jet and that of the DIS electron of data and MC. After applying this correction, the measured energy of the jets in the MC corresponds to that in the data;
    \item to improve the agreement between the detector-level and hadron-level jet energies, a further correction was applied to the detector-level jet energies in data and MC. Similar to the electron calibration described above, factors were derived from a MC study. The correction was applied as a linear function of the transverse momentum of each jet and in bins of the pseudorapidity in the laboratory frame;
    \item the agreement between data and MC distributions was improved by reweighting the hadron-level distributions of each MC sample as a function of the hadron-level jet multiplicity and average transverse momentum of the jets in each event. The corrections were derived by comparing data and MC distributions at detector level.
\end{itemize}
Distributions of the $p_{\perp,\text{Breit}}$ spectrum in different regions of $Q^2$ at detector level are shown in Fig.~\ref{fig:control jet}. The corrected \textsc{Ariadne} and \textsc{Lepto} MC distributions are compared to the data. The MC models describe the data reasonably well across the entire selected kinematic region.

\section{Cross-section determination}
\label{sec:unfolding}
In an inclusive-jet measurement, each jet that passes the selection criteria is counted individually. Consequently, events might contribute multiple times to the cross section. The inclusive-jet cross sections are defined for NC DIS events at hadron level in the kinematic region $\unit[150]{\GeV^2} < Q^2 < \unit[15000]{\GeV^2}$ and $0.2 < y < 0.7$. Hadron-level jets are identified using the same algorithm described in Section~\ref{subsec:jet selection} for detector-level jets. The jets are defined in terms of hadrons, leptons and photons with a lifetime of more than $\unit[10]{\,\text{ps}}$, excluding neutrinos. Jets were considered in the kinematic region of $-1 < \eta_\text{lab} < 2.5$ and $\unit[7]{\GeV} < p_{\perp,\text{Breit}} < \unit[50]{\GeV}$. The cross-section measurement was performed double-differentially in $Q^2$ and $p_{\perp,\text{Breit}}$. To allow direct comparison, the binning choice was taken from the corresponding H1 analysis\cite{h1highq2newjets}. It was verified that this is also a reasonable choice for the ZEUS detector.
The measured cross sections are defined for a ratio of $e^-p:e^+p$ collisions of $6:5$, corresponding to the collected luminosity.
The cross sections are defined at QED Born-level, i.e.\ at leading order in QED, but including the running of the electromagnetic coupling.

All available NLO and NNLO QCD calculations treat the underlying partons as massless.
To minimise the differences between the jet definitions at hadron and parton level, massless parton-level jets were reconstructed in the QCD calculations (from the quarks and gluons arising from the matrix elements) and the present analysis was performed using massless hadron-level jets. Correction factors were derived to make comparisons between jets constructed according to the two jet definitions possible, see Section~\ref{sec:theory}.
Cross sections for massless and massive jets differ. When treated consistently, the choice is not expected to influence conclusions drawn from the cross sections, such as the determination of $\alpha_s$.

The MC samples were used to correct the data to hadron level through two-dimensional matrix unfolding\footnote{The condition number of the migration matrix was about 12. Therefore, no regularisation was necessary during the unfolding.} as implemented in the \textsc{TUnfold} package\cite{TUnfold}. In the following, the binned two-dimensional distributions of the total number of jets in $Q^2$ and $p_{\perp,\text{Breit}}$, were mapped into one-dimensional vectors at detector level and hadron level. In the framework of matrix unfolding, the relation between detector-level and hadron-level distributions is written as
\begin{equation*}
    \Big(\mathds{1} - \diag\big(\vec{\raisebox{0pt}[7pt]{$b$}}\hspace*{1pt}\big)\Big)\cdot\vec{y} = A\cdot\vec{x},
\end{equation*}
where $\vec{x}$ is the distribution of hadron-level jets to be determined and $\vec{y}$ is the distribution of detector-level jets in the data, i.e.\ $y_i$ is the total number of observed detector-level jets in the $(Q^2,p_{\perp,\text{Breit}})$ bin indexed by $i$.
For example, if event $k$ contains two jets in one bin and a third jet in a different bin, it would have $\vec{y}_k = \big(0,\ldots,0,2,0,\ldots,0,1,0,\ldots,0\big)$. The total $\vec{y}$ is then defined as the sum over all events $\vec{y} = \sum_k \vec{y}_k$.
The matrix $A$ is the migration matrix determined from the signal MC, i.e.\ the element $A_{ij}$ represents the probability for a jet generated in hadron-level bin $j$ to be reconstructed in detector-level bin $i$. The vector $\vec{b}$ represents a generalised background fraction to be subtracted from the data before unfolding and $\diag\big(\vec{\raisebox{0pt}[7pt]{$b$}}\hspace*{1pt}\big)$ is the diagonal matrix whose diagonal elements are the entries of $\vec{b}$. Each column of $A$ represents the shape of the detector-level distribution induced by jets in the corresponding hadron-level bin. The columns of the migration matrix add up to less than one since some hadron-level jets were not reconstructed in any detector-level bin due to inefficiencies in the reconstruction or migrations out of the kinematic region.

The vector $\vec{b}$ comprises the backgrounds from events outside the kinematic range used for the unfolding and detector-level jets that cannot be assigned to any hadron-level jet in the signal MC.
The dependence on the MC models used to determine $\vec{b}$ was reduced by applying this term in a multiplicative, rather than additive, fashion. This is because this approach only required the simulation of the background MC samples to be correct relative to the signal MC, rather than relative to the data. This treatment also ensured that the absolute normalisation of the MC samples did not influence the measurement.

The unfolded distribution $\vec{x}$ was determined by minimising the expression
\begin{equation*}
    \bigg(\Big(\mathds{1} - \mathrm{diag}\big(\vec{\raisebox{0pt}[7pt]{$b$}}\hspace*{1pt}\big)\Big)\cdot\vec{y} - A\cdot\vec{x}\bigg)^\intercal
    \cdot V^{-1} \cdot
    \bigg(\Big(\mathds{1} - \mathrm{diag}\big(\vec{\raisebox{0pt}[7pt]{$b$}}\hspace*{1pt}\big)\Big)\cdot\vec{y} - A\cdot\vec{x}\bigg),
\end{equation*}
where $V$ is the covariance matrix of the measured distribution $\vec{y}$. The quantities $\vec{y}$ and $V$ were taken from the data, while $A$ and $\vec{b}$ were determined from MC samples.

The distribution of the number of inclusive jets in each bin does not follow a Poisson distribution since multiple jets can arise from the same event.
A correct treatment of statistical uncertainties and correlations can be ensured by counting $n$-jet events and assigning to them a weight $n$ rather than counting the jets themselves.
This approach was implemented using the following procedure: for each event $k$ in the data sample, an individual vector $\vec{y}_k$ was constructed similar to the vector $\vec{y}$ described above. It contained the number of jets from the event $k$. In an $n$-jet event, the vector $\vec{y}_k$ had multiple entries adding up to $n$. The vector $\vec{y}$ and the matrix $V$ were composed as $\vec{y}=\sum_k \vec{y}_k$ and $V=\sum_k \vec{y}_k \vec{y}_k{\!}^\intercal$.

To determine $A$ and $\vec{b}$ from the MC samples, jets were reconstructed at detector and hadron level and matched to each other if their separation in the $(\eta_\text{lab}-\phi)$-plane was less than $0.9$\cite{h1highq2newjets}. Matched pairs were used to fill the response matrix $R$. The element $R_{ij}$ is the number of jets generated in hadron-level bin $j$ and reconstructed in detector-level bin $i$. Unmatched hadron-level and detector-level jets were recorded in the vectors $\vec{x}_\text{miss}$ and $\vec{y}_\text{fake}$. Additionally, the detector-level distributions of the low-$Q^2$ DIS and photoproduction background MC samples were recorded in the vectors $\vec{y}_\text{Low-$Q^2$}$ and $\vec{y}_\text{PHP}$.
The migration matrix $A$ and background fraction $\vec{b}$ were then determined as
\begin{align*}
    A_{ij} &= \frac{R_{ij}}{(\sum_{i'} R_{i'j}) + x_{\text{miss},j}}, \\
    b_i &= \frac{y_{\text{fake},i}+y_{\text{Low-$Q^2$},i}+y_{\text{PHP},i}}{(\sum_j R_{ij}) + y_{\text{fake},i}+y_{\text{Low-$Q^2$},i}+y_{\text{PHP},i}}.
\end{align*}
Because of the way the MC samples were defined, jets that migrated from $\unit[100]{\GeV^2} < Q^2 < \unit[150]{\GeV^2}$ into the measurement region $Q^2 > \unit[150]{\GeV^2}$ were considered unmatched jets and contributed to $\vec{y}_\text{fake}$. Jets that migrated from $Q^2 < \unit[100]{\GeV^2}$ into the measurement region were considered low-$Q^2$ DIS background and contributed to $\vec{y}_\text{Low-$Q^2$}$. This distinction was no longer relevant during the unfolding since both contributions were treated consistently in $\vec{b}$.

To reduce the dependence on the MC model and to supply sufficient information to the unfolding procedure, most $p_{\perp,\text{Breit}}$ measurement bins were subdivided into two bins at hadron level and three bins at detector level. In $Q^2$, the measurement binning was kept at hadron level and subdivided into two at detector level. To reduce the number of missing, $\vec{x}_\text{miss}$, and background, $\vec{y}_\text{fake}$, entries in the signal region, overflow bins were added in $Q^2$ and $p_{\perp,\text{Breit}}$ up to the kinematic limit. In addition, an underflow bin in $p_{\perp,\text{Breit}}$ down to $\unit[4.5]{\GeV}$ was added. No underflow bin was added in $Q^2$, since contributions from the low-$Q^2$ DIS background sample would have become problematic in this bin. Overall, this method resulted in 63 hadron-level bins and 169 detector-level bins.

The background contribution from unmatched jets, $\vec{y}_\text{fake}$, to the detector-level distribution was about $\unit[15]{\%}$ in the central parts of the measured kinematic region and increased towards the edges. The contribution from low-$Q^2$ DIS background was less than $\unit[1]{\%}$ in most bins. Photoproduction background contributed less than $\unit[0.1]{\%}$ to the detector-level distribution.

At high $p_{\perp,\text{Breit}}$, about $\unit[30-40]{\%}$ of hadron-level jets could not be matched to any detector-level jet and thus contributed to $\vec{x}_\text{miss}$. In most cases, this was due to the corresponding event being rejected by the detector-level quality cuts or migrations in $Q^2$ or $y$. Only about $\unit[5-10]{\%}$ of hadron-level jets were unmatched because of inefficiencies in the jet reconstruction. At low $p_{\perp,\text{Breit}}$, the fraction of unmatched hadron-level jets increased to up to $\unit[60]{\%}$ due to migrations in $p_{\perp,\text{Breit}}$.

The measured cross sections were determined by unfolding the data in two different ways, using either the \textsc{Ariadne} or the \textsc{Lepto} MC model. The average of the two results was used as the nominal cross section.

As an additional check of the unfolding procedure, the analysis was repeated using a bin-by-bin acceptance correction. The two methods yielded consistent results for the cross sections and the determined value of $\alpha_s(M_Z^2)$\cite{thesis:lorkowski} (See Section~\ref{sec:fit}). This check also confirms that previous ZEUS results based on bin-by-bin acceptance corrections retain their full validity.

The uncertainties on the unfolded cross sections are correlated in $Q^2$ and $p_{\perp,\text{Breit}}$. Positive correlations in $p_{\perp,\text{Breit}}$ arise due to jets originating from the same event. Predominantly negative correlations in both quantities arise because of the finite resolution of the detector, leading to migrations between bins, as described by the migration matrix $A$. The matrix-unfolding approach considers both of these types of correlations and determines the covariance matrix of the cross sections alongside the central values.
In the following, the uncertainties determined by the unfolding procedure will be referred to as the \textit{unfolding uncertainties} $\delta_\text{unf}$. These include the statistical uncertainty from data and MC and the systematic correlations from migrations at detector level and from jets originating from the same event. The unfolding uncertainty is dominated by the statistical uncertainty on the data and also by migrations at detector level. The contribution of MC statistics is about $\unit[10]{\%}$ of the unfolding uncertainty.

For the combined QCD analysis, it is necessary to determine the correlations to the previous ZEUS dijet measurement\cite{zeusdijets}. These correlations arise since the same detector-level events were used for both measurements. Correlations were determined to the double-differential $(Q^2,\overline{p_{\perp,\text{Breit}}})$ cross sections of the dijet measurement, with $\overline{p_{\perp,\text{Breit}}}$ being the mean transverse momentum of the dijet system.
To determine the corresponding correlation matrix, the dijet event selection\cite{thesis:behr} was recreated and the dijets unfolded alongside the inclusive jets by extending the relevant vectors and matrices with additional dijet bins. Using this approach, the matrix-unfolding procedure yields the inclusive-jet correlation matrix, the dijet correlation matrix and the inclusive-jet-dijet correlation matrix.
The previous dijet measurement was performed using a bin-by-bin acceptance correction, which does not introduce correlations between the dijet points. Therefore, it is necessary to use a compatible unfolding procedure, i.e.\ the corresponding dijet correlation matrix needs to be diagonal. To ensure this, only one bin from the dijet measurement was added and unfolded at a time. The unfolding was repeated for every dijet bin. The complete inclusive-jet-dijet correlation matrix was constructed by combining the determined partial correlation matrices.
This procedure was applied to all events that were included in both measurements. Afterwards, to account for the fact that the considered run periods of the two measurements did not overlap completely, the correlations obtained were scaled by $\mathcal{L}_\text{overlap}/\sqrt{\mathcal{L}_\text{inclusive-jets} \mathcal{L}_\text{dijets}} \approx 80\%$, where $\mathcal{L}_\text{inclusive-jets}$, $\mathcal{L}_\text{dijets}$ and $\mathcal{L}_\text{overlap}$ are the integrated luminosities of the inclusive-jet measurement, the dijet measurement and the events common to both measurements, respectively.

After unfolding, the resulting hadron-level cross sections were corrected to QED Born-level, which is defined by the absence of QED-radiative effects, while including the scale dependence of the electromagnetic coupling. Corresponding MC samples were generated. Bin-wise correction factors were determined by comparing the cross sections derived from these samples to those from the nominal MC samples. These correction factors were typically in the range between $0.7$ and $0.95$, see $c_\text{QED}$ in Table~\ref{tab:cross sections}.

Cross sections are also available in an alternative definition that includes QED radiation\cite{thesis:lorkowski}. This definition allows a direct comparison to NNLO QCD + NLO electroweak theoretical predictions if such calculations become available in the future.

\section{Experimental uncertainties}
The systematic uncertainties on the measurement were estimated by changing aspects of the analysis and observing the effect on the cross sections. Instead of repeating the unfolding procedure, most uncertainties were estimated bin-by-bin by propagating the changes of the data, MC detector-level and MC hadron-level distributions to the cross sections, as this method is less susceptible to statistical fluctuations. The model uncertainty and all uncertainties evaluated using a reweighting procedure were determined using matrix unfolding.
The following sources of uncertainty were considered:
\begin{itemize}
    \item $\delta_\text{JES}$: after the corresponding correction, the remaining uncertainty in the jet-energy scale in the MC samples was estimated to be about $\unit[1]{\%}$ for jets with $p_{T,\text{lab}} > \unit[10]{\GeV}$ and about $\unit[3]{\%}$ for less energetic jets\cite{thesis:behr}. A corresponding variation of the jet energy in the MC changed the cross sections by about $\unit[4]{\%}$ at lower $Q^2$ and $\unit[2]{\%}$ at very high $Q^2$;
    \item $\delta_\text{model}$: the influence of the choice of MC model on the unfolded cross sections was estimated using a MC study. Two closure tests were performed in which each of the MC samples was, in turn, treated as pseudo-data and unfolded with the other sample. These tests are expected to reproduce the corresponding hadron-level distributions within the statistical uncertainty of the pseudo-data combined with the model uncertainty. The difference between the unfolded and hadron-level distributions was used to obtain the model uncertainty. Afterwards, the uncertainty was averaged over both closure tests, and a smoothing procedure over neighbouring bins was applied to reduce statistical fluctuations. The resulting model uncertainty was typically around $\unit[2]{\%}$ and increased to about $\unit[5]{\%}$ at the highest $Q^2$ or $p_{\perp,\text{Breit}}$;
    \item $\delta_\text{rew.}$: an alternative method was used to perform the reweighting of the MC models. In this method, each jet was individually reweighted as a function of $Q^2$ and its transverse momentum. The effect on the cross section was typically below $\unit[1.5]{\%}$;
    \item $\delta_\text{EES}$: the uncertainty on the electron-energy scale in the MC was estimated to be about $\unit[2]{\%}$\cite{thesis:behr}. A corresponding variation changed the cross sections by less than $\unit[0.5]{\%}$;
    \item $\delta_\text{EL}$: the correction of the reconstructed energy of the scattered electron was performed as a function of the azimuthal angle of the electron only, see Section~\ref{subsec:corr:jets}. The resulting change in the cross sections was typically below $\unit[1.5]{\%}$;
    \item $\delta_\text{EM}$: an alternative electron-finding algorithm was used\cite{nim:a365:508}. The effect on the cross sections was around $\unit[1]{\%}$ in most bins, with fluctuations up to $6\%$;
    \item $\delta_{p_T}$, $\delta_{E-p_\text{Z}}$, $\delta_\text{trk.}$, $\delta_\text{bal.}$, $\delta_\text{vtx.}$, $\delta_\text{rad.}$, $\delta_\text{DCA}$: the boundaries of the quality cuts on $p_{T,\text{lab}}$, $\sum(E-p_Z)$, $p_\text{track}$, $p_T/\sqrt{E_T}$, $Z_\text{vertex}$, $R_\text{RCAL}$ and the electron-track distance were varied within the resolution of each variable. The effect was typically well below $\unit[1]{\%}$, except for the $\sum(E-p_Z)$ variation, where it reached as high as $\unit[5]{\%}$ in the high-$p_{\perp,\text{Breit}}$ region;
    \item $\delta_\text{PHP}$, $\delta_\text{Low-$Q^2$}$: the backgrounds from misreconstructed photoproduction and low-$Q^2$ DIS events  ($\vec{y}_\text{PHP}$ and $\vec{y}_\text{Low-$Q^2$}$) were subtracted from the data prior to the unfolding. These distributions were taken from the MC samples and were estimated to have a $\unit[50]{\%}$ normalisation uncertainty. The resulting uncertainty on the cross sections reached about $\unit[4]{\%}$ in the lowest $Q^2$, highest $p_{\perp,\text{Breit}}$ bin and was negligible everywhere else;
    \item $\delta_\text{fake}$: similarly, the background from unmatched jets ($\vec{y}_\text{fake}$) in the signal MC was subtracted from the data prior to unfolding. From a study of the shape of the jet distribution in $Q^2$ and $p_{\perp,\text{Breit}}$, the uncertainty on the normalisation of this contribution was estimated to be $\unit[5]{\%}$. Propagating this uncertainty to the cross sections led to a systematic uncertainty of about $\unit[1.5]{\%}$ in all bins;
    \item $\delta_\text{pol.}$: the electron beam polarisation correction had an effect of less than $\unit[0.1]{\%}$ on the cross sections;
    \item $\delta_\text{TME}$: the track-association correction changed the cross sections by less than $\unit[2]{\%}$. Half of this difference was taken as the systematic uncertainty on this correction;
    \item $\delta_\text{FLT}$: the first-level-trigger veto-efficiency correction was applied as a function of the inelasticity instead of the track multiplicity. The effect on the cross sections was well below $\unit[0.5]{\%}$;
    \item $\delta_\text{QED}$: the statistical uncertainty on the MC samples used for the QED Born-level correction was added to the data as a systematic uncertainty. It was typically below $\unit[0.5]{\%}$, except for the low-$Q^2$, high-$p_{\perp,\text{Breit}}$ region, where it increased to about $\unit[3]{\%}$;
    \item to construct the response matrix, pairs of detector- and hadron-level jets had to be matched to each other. Varying the maximum allowed distance in the $(\eta_\text{lab}-\phi)$-plane from $0.9$ to $0.7$ had a negligible effect on the cross section;
    \item the uncertainty associated with the luminosity measurement was $\unit[1.9]{\%}$ for all bins. By convention, this uncertainty is not included in the figures, as it is, by definition, fully correlated across all points.
\end{itemize}
The contribution of each source to the total systematic uncertainty is shown in Fig.~\ref{fig:systematics} and listed in Tables~\ref{tab:cross sections} and \ref{tab:uncertainties}.
The overall uncertainty is dominated by the jet-energy scale and model uncertainties. In the lower $Q^2$ regions, the uncertainty of the jet-energy-scale dominates the overall uncertainty. The unfolding uncertainty becomes dominant in the highest $Q^2$ and $p_{\perp,\text{Breit}}$ bins only.

\section{Theoretical calculations}
\label{sec:theory}
Predictions for inclusive-jet production in the Breit frame are available at NNLO QCD accuracy ($\mathcal{O}(\alpha_s^3)$)~\cite{nnlojet2,gehrmann2018jet} as calculated by the \textsc{NNLOJet} program, interfaced to \textsc{fastNLO} \cite{KLUGE_2007,fastnlo3} via so-called grid files\cite{Grids,Carli_2010,Britzger_2019}. For this analysis, cross-section predictions were computed using the HERAPDF2.0Jets NNLO PDF set\cite{HERAPDF20NNLO} and using the associated value of $\alpha_s(M_Z^2) = 0.1155$. The factorisation and renormalisation scales were set to $\mu_\text{f}^2 = \mu_\text{r}^2 = Q^2+p_{\perp,\text{Breit}}^2$. The jet calculations were done in the zero-mass variable-flavour-number scheme, since calculations for massive partons are unavailable. For consistency, the constructed jets were also defined to be massless.

The parton-level jet predictions from the QCD calculations were corrected to hadron level using correction factors derived from the \textsc{Ariadne} and \textsc{Lepto} samples. For this purpose, parton-level jets were constructed in the MC samples (from the quarks, gluons and photons arising directly after the parton showering and photon radiation steps) and their ratio to hadron-level jets was computed. The average of these ratios from the two MC samples was used as the nominal correction ($c_\text{Had}$), and half their difference as hadronisation uncertainty ($\delta_\text{Had}$) on the predictions. This uncertainty reflects the differences in the corresponding parton-showering and hadronisation procedures, and is assumed to cover also the differences between the LO+PS and NNLO partons in the jet reconstruction. The size of this uncertainty is comparable to similar analyses\cite{zeusdijets,h1highq2newjets}. Since the calculations did not include weak interactions, they were also corrected for $Z$-boson exchange and $\gamma Z$-interference terms using factors derived from a separate MC sample ($c_Z$). No uncertainty was associated with this correction\cite{nnlojet2}. The correction factors are given in Table~\ref{tab:cross sections}. The calculations were performed at QED Born-level and thus correspond to the corrected cross sections described in Section~\ref{sec:unfolding}.

The uncertainty based on the variation of the factorisation and renormalisation scales was estimated using a six-point variation, in which both scales were varied up and down by a factor 2, both separately and simultaneously\cite{HERAPDF20NNLO,nnlojet2}. The fit, model and parameterisation uncertainties on the HERAPDF2.0Jets NNLO PDF set were taken into account\cite{HERAPDF20NNLO}. The grid files produced by \textsc{NNLOJet} include uncertainties due to limited statistics during their generation.

The statistical uncertainty of the grids, the factorisation- and renormalisation-scale uncertainties, the PDF uncertainties and the hadronisation uncertainty were added in quadrature to obtain the total uncertainty on the NNLO QCD predictions. In most parts of the kinematic region covered, the scale uncertainty was dominant. At high $p_{\perp,\text{Breit}}$, the parameterisation uncertainty on the PDF set was also significant.

\section{Cross-section results}
\label{sec:cross sections}
\begin{figure}[p]
    \vfill
    \begin{center}
        \includegraphics{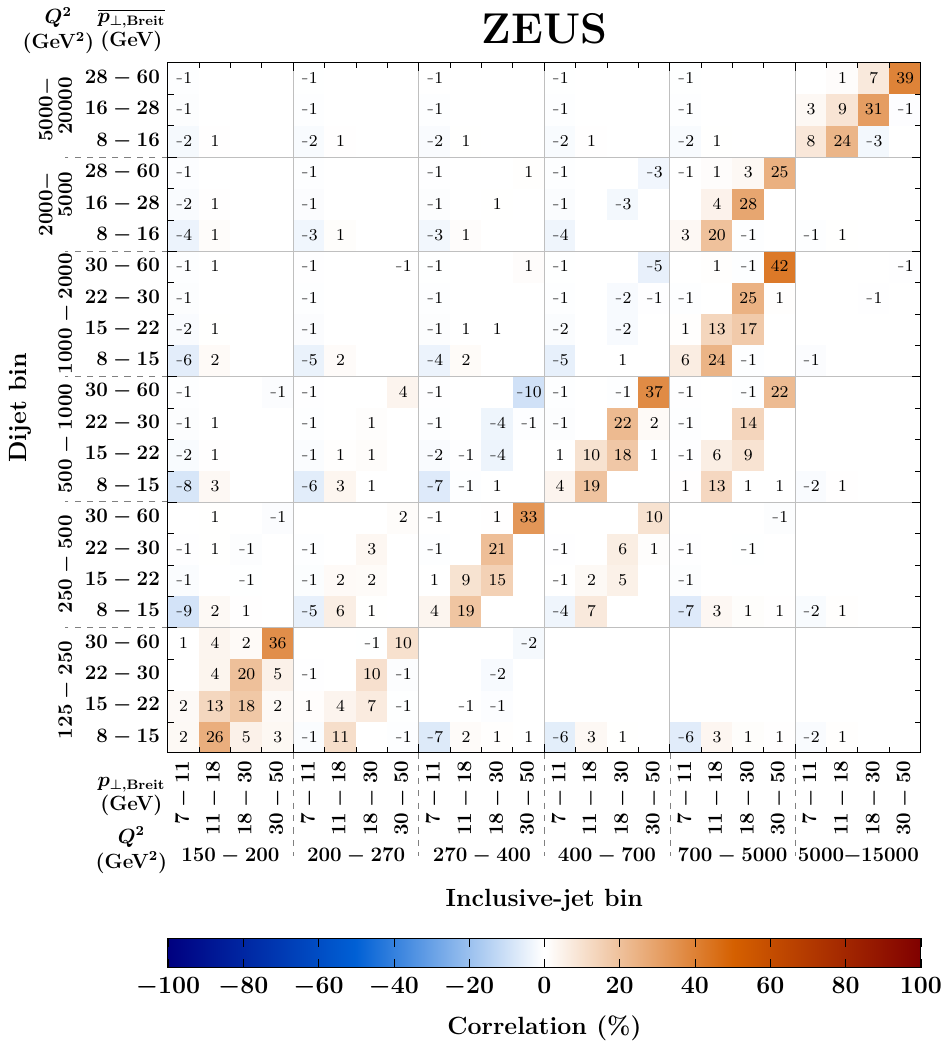}
    \end{center}
    \caption{Correlation matrix between the unfolding uncertainty of the inclusive-jet measurement and the statistical uncertainty of the previous dijet measurement\protect\cite{zeusdijets}. Correlations are mostly positive, as they arise predominantly from jets originating from the same events. A structure of more strongly correlated bins is visible, which can be explained by the differing bin boundaries of the two measurements.
    }
    \label{fig:correlation2}
    \vfill
\end{figure}

The double-differential inclusive-jet cross sections as a function of $Q^2$ and $p_{\perp,\text{Breit}}$ are shown in Fig.~\ref{fig:cross sections} and Table~\ref{tab:cross sections}. The combined uncertainty on the cross sections is typically around $\unit[5]{\%}$ and increases to around $\unit[25]{\%}$ in the highest $p_{\perp,\text{Breit}}$ bin. The correlation matrix of the inclusive-jet measurement is shown in Fig.~\ref{fig:correlation1} and listed in Table~\ref{tab:correlations}. The uncertainties entering further analysis are smaller than the uncertainties indicated by the error bars in Fig.~\ref{fig:cross sections} due to the negative correlation of the unfolding uncertainty as listed in Table~\ref{tab:correlations}. The correlations between the inclusive-jet and corresponding dijet measurement are shown in Fig.~\ref{fig:correlation2} and Table~\ref{tab:correlations}.

The corresponding measurement from the H1 collaboration\cite{h1highq2newjets} is also shown\footnote{Owing to the nature of the weak interaction, DIS cross sections involving electrons and positrons differ at high $Q^2$. The cross sections represent a luminosity-weighted average of the $e^+p$ and $e^-p$ data. The compositions of the ZEUS and H1 data differ slightly. If the H1 cross sections were corrected to the ZEUS composition, they would increase by about $\unit[1]{\%}$ in the fourth and fifth $Q^2$ bin and by about $\unit[5]{\%}$ in the highest $Q^2$ bin. In Fig.~\ref{fig:cross sections}, the values are shown as published by H1.} in Fig.~\ref{fig:cross sections}.
The H1 measurement agrees very well with the ZEUS cross section and the uncertainties are comparable. Both measurements show similar trends relative to the NNLO QCD predictions. Within the combined uncertainty, the NNLO QCD predictions agree reasonably well with the measured cross sections. Overall, the central values of the predictions seem to overestimate the jet cross section. At high $p_{\perp,\text{Breit}}$, this difference increases.

\section{Determination of the strong coupling constant}
\label{sec:fit}
Predictions of jet cross sections depend, among other ingredients, on the PDFs and the strong coupling constant, $\alpha_s(M_Z^2)$. Since the former belong to the realm of non-perturbative QCD, they cannot currently be calculated from first principles, but only obtained from fits to data. The double-differential inclusive-jet cross sections are particularly well suited to constrain these fits because of their direct sensitivity to $\alpha_s(M_Z^2)$ and their small experimental and theoretical uncertainties. The measured cross sections were used as input to a QCD analysis at NLO and NNLO to perform a simultaneous determination of the PDFs of the proton and the strong coupling constant.

The inclusion of jet data in the fit is expected to reduce strongly the dependence of the measured strong coupling constant on the assumed gluon distribution in the proton. To capture these correlations, a simultaneous fit of $\alpha_s(M_Z^2)$ and the PDFs was performed. The PDFs were parameterised using the HERAPDF ansatz\cite{HERAPDF20NNLO}. The input to the fit consisted of the H1+ZEUS combined inclusive DIS dataset\cite{HERAPDF20NLO}, previous inclusive-jet\cite{pl:b547:164} and dijet\cite{zeusdijets,Currie_2016} measurements at ZEUS and the inclusive-jet cross sections of this paper. Because of a cut on the invariant mass in the dijet measurement, the $\mathcal{O}(\alpha_s)$ prediction vanishes for parts of the dijet phase space, which leads to an increased scale uncertainty in the corresponding fixed-order calculations. To avoid this issue, six dijet points at low $p_{\perp,\text{Breit}}$ were excluded from the analysis\cite{HERAPDF20NNLO}.

Statistical and systematic correlations between the dijet measurement and the present inclusive-jet measurement were taken into account. Statistical correlations arise since the same detector-level events were used for both measurements and were treated using a correlation matrix as described in Section~\ref{sec:unfolding}.
Systematic correlations arise because the jet-energy-scale and the luminosity uncertainties have a similar effect on both measurements. These sources were treated as \unit[80]{\%} correlated between the two measurements. This is due to the overlap in data samples, which was described in Section~\ref{sec:unfolding}. The present and the previous\cite{pl:b547:164} inclusive-jet measurements were treated as uncorrelated.

The uncertainties associated with the relative normalisation of the background from low-$Q^2$ DIS events ($\delta_\text{Low-$Q^2$}$), the $(E-p_\text{Z})$-cut boundaries ($\delta_{E-p_\text{Z}}$) and the track-matching-efficiency correction ($\delta_\text{TME}$) were treated as fully correlated within the inclusive-jet dataset. The uncertainties associated with the choice of the MC model ($\delta_\text{model}$) and the relative normalisation of the background from unmatched detector-level jets ($\delta_\text{fake}$) were treated as half-correlated and half-uncorrelated. All remaining uncertainties were added in quadrature and treated as uncorrelated ($\delta_\text{uncor}$). For use in the fit, all uncorrelated uncertainties and the jet-energy-scale uncertainty were symmetrised by averaging their positive and negative components.

The fit was performed similarly to HERAPDF analyses\cite{HERAPDF20NLO,HERAPDF20NNLO}. The following parameters were used in the nominal fit and the stated variations were used to determine uncertainties. Inclusive DIS data points were constrained by requiring $Q^2 \geq Q^2_\text{min} = \unit[3.5 \begin{smallmatrix}+1.5\\-1.0\end{smallmatrix}]{\GeV^2}$. The starting scale for the DGLAP evolution was set to $\mu_{\text{f}0}^2 = \unit[1.9 \pm 0.3]{\GeV^2}$. Heavy-quark masses in the calculations of the inclusive DIS cross sections were set to $m_c = \unit[1.46 \pm 0.04]{\GeV}$ and $m_b = \unit[4.3 \pm 0.10]{\GeV}$ at NLO, and $m_c = \unit[1.41 \pm 0.04]{\GeV}$ and $m_b = \unit[4.2 \pm 0.10]{\GeV}$ at NNLO. The strange-quark content of the down-type sea was set to $f_s=0.4 \pm 0.1$. In the jet calculations, the factorisation and renormalisation scales were set to $\mu_\text{f}^2=Q^2$ and $\mu_\text{r}^2=(Q^2+p_\perp^2)/2$ at NLO\cite{HERAPDF20NLO} and $\mu_\text{f}^2=\mu_\text{r}^2=Q^2+p_\perp^2$ at NNLO\cite{HERAPDF20NNLO}, where $p_\perp$ is $p_{\perp,\text{Breit}}$ for the inclusive-jet calculations and $\overline{p_{\perp,\text{Breit}}}$ for the dijet calculations. The fit was performed using the \textsc{xFitter} program\cite{HERAFitter,xFitter,JAMES1975343,Botje_2011,Bertone_2014}.

Using the standard scheme of fully correlated scale variations, the fit resulted in values of
\begin{align*}
    \text{NNLO:~} \alpha_s(M_Z^2) &= 0.1143
        \pm 0.0017 \text{~\small(exp./fit)~} 
        \unc{06}{07} \text{\small~(model/param.)~}
        \unc{12}{05} \text{\small~(scale)}, \\
    \text{NLO:~} \alpha_s(M_Z^2) &= 0.1160
        \pm 0.0017 \text{\small~(exp./fit)~}
        \unc{07}{09} \text{\small~(model/param.)~}
        \unc{26}{14} \text{\small~(scale)},
\end{align*}
where `exp./fit' denotes the uncertainty on the fit, which includes the uncertainty in the experimental input together with that of the hadronisation correction and the statistical uncertainty on the NNLO grids. The additional model and parameterisation uncertainty was determined by repeating the fit with each of the input quantities listed above in turn modified by their uncertainty and adding the resulting variations of $\alpha_s(M_Z^2)$ in quadrature, separately for positive and negative uncertainties. To ensure that the starting scale stayed below the heavy-quark masses, the variation of $\mu_{\text{f}0}^2$ was performed only downward and the variation of $m_\text{c}$ only upward and the resulting change of $\alpha_s(M_Z^2)$ was symmetrised. For each of the eight HERAPDF $D$ and $E$ parameters that were not considered in the nominal fit, an additional fit was performed in which one more parameter was left free\cite{HERAPDF20NNLO}. The envelope of the resulting $\alpha_s(M_Z^2)$ values was taken as the second contribution to the model/parameterisation uncertainty.
The scale uncertainty was evaluated by performing six additional fits, in which the factorisation and renormalisation scales, $\mu_{\text{f}}$ and $\mu_{\text{r}}$, for the jet cross sections were varied by a factor $2$ and taking the envelope of the resulting $\alpha_s(M_Z^2)$ values.
The nominal fits obtained a $\chi^2$ per degree of freedom of $1419/1200$ at NNLO and $1415/1200$ at NLO. The partial $\chi^2$ values at NNLO are given in table~\ref{tab:chi2}. The jet data are fully consistent in the inclusive DIS data and they reduce the value of $\chi^2$ per degree of freedom.

The scale uncertainties obtained here are significantly smaller than those derived in similar determinations, e.g.\ the HERAPDF analysis\cite{HERAPDF20NNLO}. This arises mostly because in this analysis only jet datasets at high $Q^2$ were used.
Owing to the treatment of the cross-section scale uncertainty as fully correlated across all phase-space regions, the inclusion of low-$Q^2$ data leads to an increased uncertainty on $\alpha_s(M_Z^2)$.

An alternative approach for the treatment of scale uncertainties was investigated. In this case, the scale uncertainties on the jet contribution\footnote{The fixed scale $\mu^2 = Q^2$ for the inclusive DIS cross sections was treated as part of the PDF definition.} were calculated under the assumption that the cross-section uncertainty due to the scale variation was half-correlated and half-uncorrelated between bins and datasets. This was motivated by the fact that, while the scale dependence of neighbouring phase-space bins is certainly very strongly correlated, the scale dependence of bins far away from each other in phase space, or for different final states, can be much less correlated or even anti-correlated. Such a half-correlated and half-uncorrelated approach has been used in previous analyses\cite{h1highq2newjets,HERAPDF20NLO}.

For the uncorrelated contribution, the scale uncertainty on the cross section predictions was evaluated as described in Section~\ref{sec:theory} using the PDFs and $\alpha_s(M_Z^2)$ value from the nominal fit. These uncertainties were scaled down by a factor $\sqrt{2}$ and added to the fit as uncorrelated relative uncertainties. The central value from this fit was used as the central value of this alternative $\alpha_s(M_Z^2)$ determination and the resulting increase of the fit uncertainty on $\alpha_s(M_Z^2)$ was treated as the uncorrelated contribution to the scale uncertainty.
For the correlated contribution, six additional fits were performed corresponding to a six-point variation of the factorisation and renormalisation scales with rescaling factors $\sqrt{0.5}$ and $\sqrt{2}$. The envelope of the resulting $\alpha_s(M_Z^2)$ values was taken as the correlated uncertainty. The complete scale uncertainty was obtained by adding the uncorrelated and the correlated contributions in quadrature.

Using this approach resulted in values of
\begin{align*}
    \text{NNLO:~} \alpha_s(M_Z^2) &= 0.1142
        \pm 0.0017 \text{~\small(exp./fit)~} 
        \unc{06}{07} \text{\small~(model/param.)~}
        \unc{06}{04} \text{\small~(scale)}, \\
    \text{NLO:~} \alpha_s(M_Z^2) &= 0.1159
        \pm 0.0017 \text{\small~(exp./fit)~}
        \unc{07}{09} \text{\small~(model/param.)~}
        \unc{12}{09} \text{\small~(scale)}.
\end{align*}
Even though the fit does not contain any low-$Q^2$ jet data, the reduction in the scale uncertainty is large, both at NNLO and NLO. The half-correlated and half-uncorrelated approach is expected to have an even more significant impact when using input data across a wider range in phase space.

\begin{figure}[p]
    \vfill
    \begin{center}
        \includegraphics{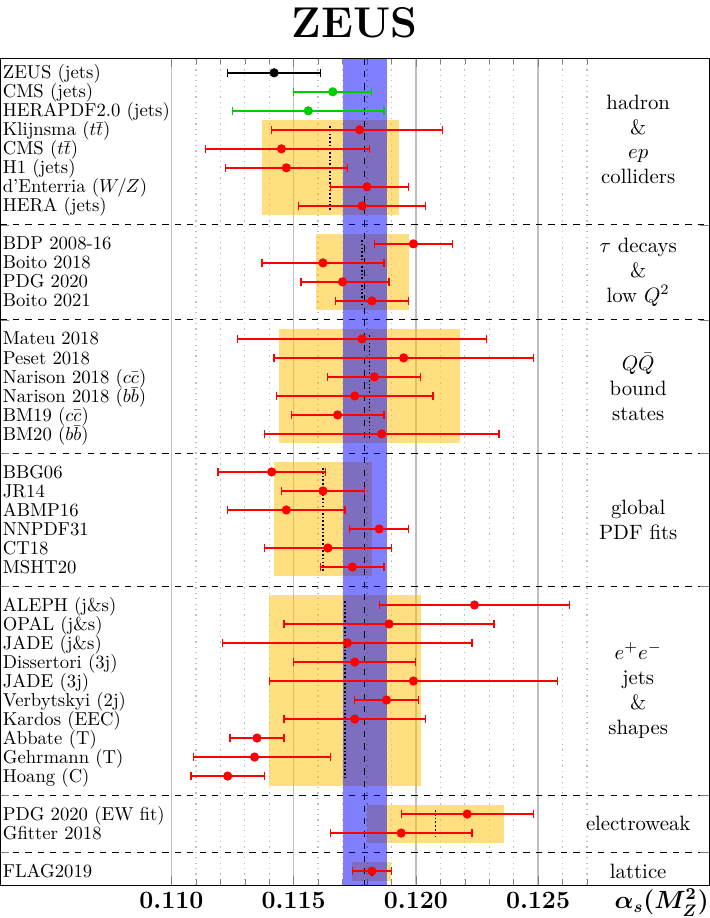}
    \end{center}
    \caption{Summary of different determinations of $\alpha_s(M_Z^2)$ at NNLO or higher order, adapted from PDG\protect\cite{10.1093/ptep/ptac097}, see references therein. The red points are included in the PDG world average. The averages from each sub-field are shown as yellow bands and the world average as a blue band. A recent measurement from CMS\protect\cite{CMSjets} using jet cross sections and the latest determination from HERAPDF\protect\cite{HERAPDF20NNLO}, which are not yet included in the world average, are shown in green. The current determination, assuming half-correlated and half-uncorrelated scale uncertainties, is shown in black.}
    \label{fig:strong coupling}
    \vfill
\end{figure}

A comparison of the current measurement to other determinations of $\alpha_s(M_Z^2)$ is shown in Fig.~\ref{fig:strong coupling}. The current analysis is among the most precise measurements at colliders.

The values determined in the fit with free $\alpha_s(M_Z^2)$ were confirmed by performing a $\chi^2$ scan with fixed $\alpha_s(M_Z^2)$ values. The results are in excellent agreement. The $\chi^2$ values of the scan at NNLO are depicted in Fig.~\ref{fig:scan}. Fits were also performed using only the previous ZEUS jet datasets or using only the newly measured dataset\cite{thesis:lorkowski}. The results were found to be consistent with the values reported here.

The calculated cross-section values before and after including the inclusive-jet dataset in the fit are compared to the data in Fig.~\ref{fig:fit nnlo}. The changes of the PDFs and $\alpha_s(M_Z^2)$ through the inclusion of the additional jet data decreased the resulting cross-section values slightly. At large $p_{\perp,\text{Breit}}$, the effect is more pronounced. The largest contribution comes from the updated value of $\alpha_s(M_Z^2)$ as well as the gluon PDF. The quark PDFs were not significantly affected by the inclusion of the additional data.

\section{Running of the strong coupling}
\label{sec:running}
A further analysis has been performed to demonstrate the scale dependence of $\alpha_s(\mu^2)$. The approach is conceptually different from the global determination. Only subsets of the measured jet cross sections were used, each centred around a certain value of the scale $\langle\mu\rangle$. Since the PDFs cannot be usefully constrained from the jet data at one scale only, it is not feasible to fit the PDFs and $\alpha_s$ at the same time. Instead, fixed PDFs were used, which were determined from the inclusive DIS data alone for different $\alpha_s(M_Z^2)$ values.
Using these PDFs, a single-parameter fit of the strong coupling was performed. While the technical fit parameter was still $\alpha_s(M_Z^2)$, this fit effectively determined $\alpha_s(\langle\mu\rangle^2)$ since only data at the scale $\langle\mu\rangle$ were used as input\cite{thesis:lorkowski}. The value of $\alpha_s(\langle\mu\rangle^2)$ was computed from $\alpha_s(M_Z^2)$ using QCD evolution. Such a procedure correctly determines $\alpha_s(\langle\mu\rangle^2)$, even if the true scale dependence of $\alpha_s$ was different from the QCD prediction.

This approach reduces the ability of the jet data to constrain the shape of the PDFs and assumes that they do so only via correlations to $\alpha_s(M_Z^2)$.
This is justified since a recent HERAPDF analysis demonstrated that the impact of the jet data on the PDFs was small\cite{HERAPDF20NNLO}.

The analysis was performed at NNLO. PDFs were determined from the inclusive DIS data, using fixed values of $\alpha_s(M_Z^2)$ between $0.112$ and $0.120$. Central values of the PDFs were determined including experimental, model and parameterisation uncertainties similar to those of the HERAPDF2.0 NNLO analysis\cite{HERAPDF20NLO}.

Each of the jet cross sections from the three datasets specified in the previous section was assigned a scale using an approximation of the barycentre of the corresponding bin,
\begin{align*}
    \frac{1}{\mu^4} &= \frac{1}{2}\left(\frac{1}{(Q^2_\text{low} + p^2_{\perp,\text{low}})^2} + \frac{1}{(Q^2_\text{high} + p^2_{\perp,\text{high}})^2}\right),
\end{align*}
where $(Q^2/p_\perp)_\text{low/high}$ are the lower/upper bin boundaries and $p_\perp$ is $p_{\perp,\text{Breit}}$ for the inclusive jets and $\overline{p_{\perp,\text{Breit}}}$ for the dijets. The resulting scales $\mu$ cover a range from $\unit[15]{\GeV}$ to $\unit[90]{\GeV}$. The points were then sorted into five different groups of similar scale. Each group was assigned a representative scale value $\langle\mu\rangle$ by computing the cross-section-weighted average of the scale values of the data points in that group.

The value of $\alpha_s(\langle\mu^2\rangle)$ was extracted for each scale $\langle\mu\rangle$ by using the jet cross sections in the respective group. Technically, \textsc{xFitter} always uses $\alpha_s(M_Z^2)$ as a parameter for $\alpha_s$. Therefore, a $\chi^2$ scan in $\alpha_s(M_Z^2)$ was performed for each group, i.e.\ the $\chi^2(\alpha_s(M_Z^2))$ values were computed for a series of fixed values of $\alpha_s(M_Z^2)$ and the corresponding fixed PDFs determined as described above. This minimised the impact of the inclusive data which contributed only indirectly via the PDFs. The $\chi^2$-definition was similar to that used in HERAPDF\cite{HERAPDF20NLO} and included the uncertainties of the PDFs. The $\chi^2(\alpha_s(M_Z^2))$ dependence close to its minimum was fitted with a parabola. The central $\alpha_s(M_Z^2)$ value and its uncertainty were extracted from the location of the minimum of the parabola and its width at the height where it has increased by one unit in $\chi^2$ with respect to its minimum. The values of $\alpha_s(\langle\mu^2\rangle)$ were then calculated from the central values $\alpha_s(M_Z^2)$ using NNLO QCD evolution.

This method intrinsically provides the sum of the experi\-mental/fit and the model/\allowbreak{}para\-meter\-isation uncertainties. To separate them, the model/para\-meter\-isation uncertainty of the PDF set was set to zero and the determination was repeated. It was found that the central value of $\alpha_s$ did not change significantly and, as expected, the uncertainty decreased. The uncertainty from this second determination was taken as the experi\-mental/fit uncertainty and the quadratic difference from the full uncertainty was taken as the model/para\-meter\-isation uncertainty.

The scale uncertainty was determined by repeating the determination six more times corresponding to a six-point variation of the factorisation and renormalisation scales with rescaling factors $0.5$ and $2$. The envelope of the resulting $\alpha_s(M_Z^2)$ values was taken as the scale uncertainty. This assumes that the scale dependence of the cross section is fully correlated across all jet cross sections in a particular group. This assumption is appropriate here since this determination used jet cross sections in a much smaller part of the kinematic region than was used for the global determination.

As a cross check, the same procedure was also applied to all the jet cross sections simultaneously. The determined value of the strong coupling constant is $\alpha_s(M_Z^2) = 0.1161 \pm 0.0019 \text{\small~(exp./fit)}$ $\pm 0.0004 \text{\small~(model/param.)}$ $\unc{14}{07} \text{\small~(scale)}$. This value is slightly different from that found in the global determination and has a slightly larger uncertainty.
This is expected because only the inclusive DIS data were used in the pre-determinations of the $\alpha_s(M_Z^2)$-dependent PDFs and only the jet data were used in the fits to extract $\alpha_s(\langle\mu\rangle^2)$ from $\chi^2$ scans. Thus, the cross-correlations are not treated as comprehensively as in the combined fit.

\begin{figure}[p]
    \vfill
    \begin{center}
        \includegraphics{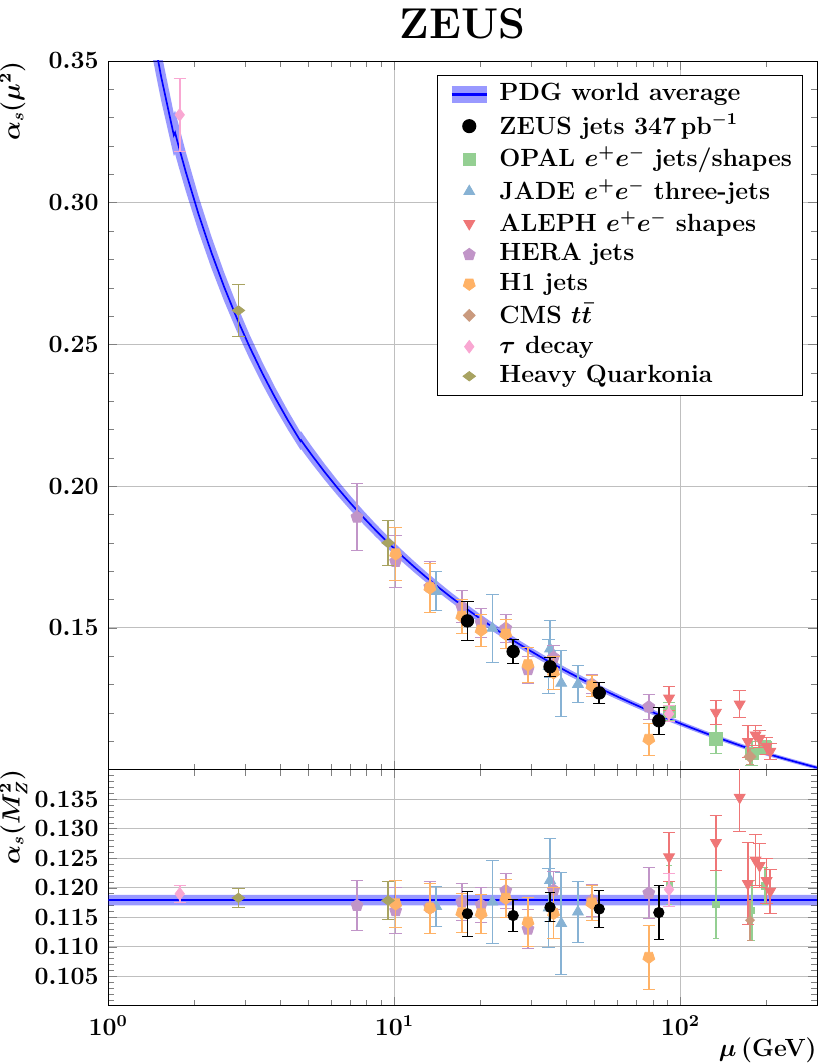}
    \end{center}
    \caption{Value of the strong coupling $\alpha_s(\mu^2)$ as a function of the scale $\mu$. The data points indicate determinations from measurements that were performed close to the indicated scale. The uncertainties represent the full uncertainty of each determination. All depicted results were obtained at least at NNLO. They are based on data from $e^+e^-$\protect\cite{Abbiendi_2011,Schieck_2013,Dissertori_2008}, $ep$\protect\cite{Britzger_2019,H1alphas} and $pp$\protect\cite{CMSalphas} collisions, as well as from $\tau$ lepton decays\protect\cite{Pich_2014} and quarkonium states\protect\cite{Narison_2018}. The solid blue line shows the PDG world average\protect\cite{10.1093/ptep/ptac097}. Also shown are the $\alpha_s(M_Z^2)$ values corresponding to each data point.
    }
    \label{fig:running}
    \vfill
\end{figure}

The determined values of the strong coupling are given in Table~\ref{tab:running} at the $Z$-boson mass and at the scale of each group of cross sections. All five values are very well compatible with the result of the global determination. Previous measurements and the QCD prediction of the running of $\alpha_s(\mu^2)$ are compared to the data in Fig.~\ref{fig:running}. The measurements are consistent with each other and with the theoretical expectation.
This confirms that the scale evolution of $\alpha_s$ can be described with the standard formalism of perturbative QCD.

\section{Summary and conclusions}
A measurement of the double-differential inclusive-jet cross section in the Breit frame in NC DIS events has been presented. The data entering the analysis were taken with the ZEUS detector at HERA between the years 2004 and 2007 at a centre-of-mass energy of $\unit[318]{\GeV}$ and correspond to an integrated luminosity of $\unit[347]{\pbi}$. The massless jets were reconstructed using the $k_t$-algorithm in the Breit frame of reference in the range $\unit[7]{\GeV} < p_{\perp,\text{Breit}} < \unit[50]{\GeV}$ and $-1 < \eta_\text{lab} < 2.5$. The cross sections were measured in the DIS kinematic region $\unit[150]{\GeV^2} < Q^2 < \unit[15000]{\GeV^2}$ and $0.2 < y < 0.7$.
The uncertainties on the measured cross sections are comparable to previous measurements. Within uncertainties, previous measurements and NNLO QCD predictions agree well with these cross sections.
The cross section data are available in a machine-readable format on \textsc{HEPdata}~\cite{hepdata.145637}. Data files for the \textsc{xFitter} program are available in the official repository~\cite{xFitter}.

The small uncertainties on the cross sections and the corresponding theory calculations make the dataset well suited for precision determinations of the strong coupling in QCD fits. A significant reduction of the scale uncertainties with respect to previous determinations was found to be predominantly due to the absence of low-$Q^2$ jet data in the fit. An improved treatment of the correlations in the scale uncertainties further reduced the uncertainties.
The value of the strong coupling constant at NNLO was determined to be $\alpha_s(M_Z^2) = 0.1142 \pm 0.0017 \text{~\small(exp./fit)}$ $\unc{06}{07} \text{\small~(model/parameterisation)}$ $\unc{06}{04} \text{\small~(scale)}$.
The dependence of the strong coupling on the energy scale was found to be consistent with previous measurements and the perturbative QCD expectation.

\section*{Acknowledgements}
\label{sec-ack}

We appreciate the contributions to the construction, maintenance and operation of the ZEUS detector of many people who are not listed as authors. The HERA machine group and the DESY computing staff are especially acknowledged for their success in providing excellent operation of the collider and the data-analysis environment. We thank the DESY directorate for their strong support and encouragement.
We thank Patrick Connor for many useful discussions concerning the matrix unfolding of the inclusive-jet cross sections.

\clearpage

\pdfbookmark[section]{Bibliography}{bib}
\bibliographystyle{zeus_bib_style}
{\raggedright\bibliography{InclusiveJets.bib}}
\vfill\eject

\let\oldtext\text
\renewcommand{\text}[1]{\oldtext{\upshape#1}}
\begin{table}[p]
    \pdfbookmark[section]{Tables}{tab}
    \vfill
    \centering
    \scalebox{0.8}{\begin{tabular}{|@{\,}>{\boldmath\bfseries}r>{\boldmath\bfseries}r|r@{.}lrrrrrrrr|r|rrr|} \hline
        \multicolumn{1}{|c}{\hspace*{-0.1cm}\boldmath\bfseries\makecell{$Q^2$                   \\$(\text{GeV}^2)$}\hspace*{-0.1cm}} &
        \multicolumn{1}{c|}{\hspace*{-0.5cm}\boldmath\bfseries\makecell{$p_{\perp,\text{Breit}}$\\$(\text{GeV})$  }\hspace*{-0.1cm}} &
        \multicolumn{2}{c }{\hspace*{-0.1cm}\boldmath\bfseries\makecell{$\sigma$                \\$(\text{pb})$   }\hspace*{-0.1cm}} &
        \multicolumn{1}{c }{\hspace*{-0.1cm}\boldmath\bfseries\makecell{$\delta_\text{unf}$\!   \\$(\%)$          }\hspace*{-0.1cm}} &
        \multicolumn{1}{c }{\hspace*{-0.1cm}\boldmath\bfseries\makecell{\!$\delta_\text{uncor}$\!\\$(\%)$         }\hspace*{-0.1cm}} &
        \multicolumn{1}{c }{\hspace*{-0.1cm}\boldmath\bfseries\makecell{\!$\delta_\text{JES}$\! \\$(\%)$          }\hspace*{-0.1cm}} &
        \multicolumn{1}{c }{\hspace*{-0.1cm}\boldmath\bfseries\makecell{\!$\delta_\text{model}$\!\\$(\%)$         }\hspace*{-0.1cm}} &
        \multicolumn{1}{c }{\hspace*{-0.1cm}\boldmath\bfseries\makecell{\!$\delta_\text{fake}$\!\\$(\%)$          }\hspace*{-0.1cm}} &
        \multicolumn{1}{c }{\hspace*{-0.1cm}\boldmath\bfseries\makecell{\!$\delta_\text{Low-$Q^2$}$\!\\$(\%)$     }\hspace*{-0.1cm}} &
        \multicolumn{1}{c }{\hspace*{-0.1cm}\boldmath\bfseries\makecell{\!$\delta_{E-p_\text{z}}$\! \\$(\%)$      }\hspace*{-0.1cm}} &
        \multicolumn{1}{c|}{\hspace*{-0.1cm}\boldmath\bfseries\makecell{\!$\delta_\text{TME}$   \\$(\%)$          }\hspace*{-0.1cm}} &
        \multicolumn{1}{c|}{\hspace*{-0.1cm}\boldmath\bfseries\makecell{$c_\text{QED}$                            }\hspace*{-0.1cm}} &
        \multicolumn{1}{c }{\hspace*{-0.1cm}\boldmath\bfseries\makecell{$c_{Z}$                                   }\hspace*{-0.1cm}} &
        \multicolumn{1}{c }{\hspace*{-0.1cm}\boldmath\bfseries\makecell{$c_\text{Had}$                            }\hspace*{-0.1cm}} &
        \multicolumn{1}{c|}{\hspace*{-0.1cm}\boldmath\bfseries\makecell{$\delta_\text{Had}$     \\$(\%)$          }\hspace*{-0.1cm}} \\ \hline
        \BeginAccSupp{ActualText=0150-00200 07-11 68.4265 03.62 1.67 +4.41-4.07 01.99 1.72 +0.29 +0.89-0.01 1.41 0.686 1.00 0.93 0.8}    150--200 &  7--11 & 68&4   &  3.6 & 1.7 & ${}^{+4.4}_{-4.1}$ &  2.0 & 1.7 &  0.3 & ${}^{+0.9}_{-0.0}$ & 1.4 & 0.69 & 1.00 & 0.93 & 0.8 \EndAccSupp{} \\
        \BeginAccSupp{ActualText=0150-00200 11-18 29.5945 04.46 1.76 +5.27-5.09 00.97 2.11 +1.13 +0.80-0.23 1.39 0.868 1.00 0.97 0.9}    150--200 & 11--18 & 29&6   &  4.5 & 1.8 & ${}^{+5.3}_{-5.1}$ &  1.0 & 2.1 &  1.1 & ${}^{+0.8}_{-0.2}$ & 1.4 & 0.87 & 1.00 & 0.97 & 0.9 \EndAccSupp{} \\
        \BeginAccSupp{ActualText=0150-00200 18-30 06.5466 08.29 1.68 +3.39-3.22 01.35 1.19 +1.74 +1.79-0.32 1.35 0.946 1.00 0.97 0.4}    150--200 & 18--30 &  6&55  &  8.3 & 1.7 & ${}^{+3.4}_{-3.2}$ &  1.4 & 1.2 &  1.7 & ${}^{+1.8}_{-0.3}$ & 1.3 & 0.95 & 1.00 & 0.97 & 0.4 \EndAccSupp{} \\
        \BeginAccSupp{ActualText=0150-00200 30-50 00.8279 22.43 5.34 +3.41-4.15 04.16 1.05 +4.23 +5.34+0.04 1.24 0.901 1.00 0.96 0.3}    150--200 & 30--50 &  0&828 & 22.4 & 5.3 & ${}^{+3.4}_{-4.2}$ &  4.2 & 1.1 &  4.2 & ${}^{+5.3}_{+0.0}$ & 1.2 & 0.90 & 1.00 & 0.96 & 0.3 \EndAccSupp{} \\ \hline
        \BeginAccSupp{ActualText=0200-00270 07-11 55.8584 03.70 1.60 +4.10-3.87 02.27 1.29 +0.21 +0.66-0.18 1.00 0.679 1.00 0.94 1.2}    200--270 &  7--11 & 55&9   &  3.7 & 1.6 & ${}^{+4.1}_{-3.9}$ &  2.3 & 1.3 &  0.2 & ${}^{+0.7}_{-0.2}$ & 1.0 & 0.68 & 1.00 & 0.94 & 1.2 \EndAccSupp{} \\
        \BeginAccSupp{ActualText=0200-00270 11-18 27.8190 04.30 1.32 +4.71-4.54 00.95 1.55 +0.14 +1.06-0.15 0.99 0.800 1.00 0.96 0.2}    200--270 & 11--18 & 27&8   &  4.3 & 1.3 & ${}^{+4.7}_{-4.5}$ &  0.9 & 1.5 &  0.1 & ${}^{+1.1}_{-0.2}$ & 1.0 & 0.80 & 1.00 & 0.96 & 0.2 \EndAccSupp{} \\
        \BeginAccSupp{ActualText=0200-00270 18-30 06.7032 07.84 2.79 +3.72-3.36 -2.00 0.97 -0.10 +2.04-0.33 0.96 0.953 1.00 0.98 0.2}    200--270 & 18--30 &  6&70  &  7.8 & 2.8 & ${}^{+3.7}_{-3.4}$ & -2.0 & 1.0 & -0.1 & ${}^{+2.0}_{-0.3}$ & 1.0 & 0.95 & 1.00 & 0.98 & 0.2 \EndAccSupp{} \\
        \BeginAccSupp{ActualText=0200-00270 30-50 00.5901 28.89 7.06 +3.63-4.08 05.55 1.37 -2.33 +4.52-1.20 0.90 0.951 1.00 0.98 0.8}    200--270 & 30--50 &  0&590 & 28.9 & 7.1 & ${}^{+3.6}_{-4.1}$ &  5.5 & 1.4 & -2.3 & ${}^{+4.5}_{-1.2}$ & 0.9 & 0.95 & 1.00 & 0.98 & 0.8 \EndAccSupp{} \\ \hline
        \BeginAccSupp{ActualText=0270-00400 07-11 51.2466 03.67 1.69 +3.86-3.56 02.54 1.33 +0.07 +0.79+0.06 0.75 0.704 1.00 0.93 1.6}    270--400 &  7--11 & 51&2   &  3.7 & 1.7 & ${}^{+3.9}_{-3.6}$ &  2.5 & 1.3 &  0.1 & ${}^{+0.8}_{+0.1}$ & 0.7 & 0.70 & 1.00 & 0.93 & 1.6 \EndAccSupp{} \\
        \BeginAccSupp{ActualText=0270-00400 11-18 26.5737 04.28 2.29 +4.19-4.26 00.94 1.37 +0.16 +1.03-0.01 0.75 0.690 1.00 0.97 0.1}    270--400 & 11--18 & 26&6   &  4.3 & 2.3 & ${}^{+4.2}_{-4.3}$ &  0.9 & 1.4 &  0.2 & ${}^{+1.0}_{-0.0}$ & 0.8 & 0.69 & 1.00 & 0.97 & 0.1 \EndAccSupp{} \\
        \BeginAccSupp{ActualText=0270-00400 18-30 06.5587 07.17 2.19 +3.56-3.74 01.72 1.00 +0.18 +1.38-0.50 0.72 0.923 1.00 0.97 0.2}    270--400 & 18--30 &  6&56  &  7.2 & 2.2 & ${}^{+3.6}_{-3.7}$ &  1.7 & 1.0 &  0.2 & ${}^{+1.4}_{-0.5}$ & 0.7 & 0.92 & 1.00 & 0.97 & 0.2 \EndAccSupp{} \\
        \BeginAccSupp{ActualText=0270-00400 30-50 00.6898 24.02 4.96 +3.77-2.92 -5.25 1.08 +1.03 +4.99-1.60 0.67 0.956 1.00 0.98 0.4}    270--400 & 30--50 &  0&690 & 24.0 & 5.0 & ${}^{+3.8}_{-2.9}$ & -5.2 & 1.1 &  1.0 & ${}^{+5.0}_{-1.6}$ & 0.7 & 0.96 & 1.00 & 0.98 & 0.4 \EndAccSupp{} \\ \hline
        \BeginAccSupp{ActualText=0400-00700 07-11 45.5135 03.55 1.07 +3.34-3.03 02.41 1.18 +0.17 +0.89-0.23 0.58 0.755 1.01 0.92 1.4}    400--700 &  7--11 & 45&5   &  3.5 & 1.1 & ${}^{+3.3}_{-3.0}$ &  2.4 & 1.2 &  0.2 & ${}^{+0.9}_{-0.2}$ & 0.6 & 0.75 & 1.01 & 0.92 & 1.4 \EndAccSupp{} \\
        \BeginAccSupp{ActualText=0400-00700 11-18 26.6598 03.44 1.17 +3.52-3.56 01.82 1.19 -0.06 +1.71-0.05 0.59 0.626 1.01 0.97 0.2}    400--700 & 11--18 & 26&7   &  3.4 & 1.2 & ${}^{+3.5}_{-3.6}$ &  1.8 & 1.2 & -0.1 & ${}^{+1.7}_{-0.0}$ & 0.6 & 0.63 & 1.01 & 0.97 & 0.2 \EndAccSupp{} \\
        \BeginAccSupp{ActualText=0400-00700 18-30 08.2323 05.65 1.04 +3.50-3.35 02.18 0.90 -0.01 +1.73-0.24 0.59 0.854 1.01 0.97 0.3}    400--700 & 18--30 &  8&23  &  5.6 & 1.0 & ${}^{+3.5}_{-3.3}$ &  2.2 & 0.9 & -0.0 & ${}^{+1.7}_{-0.2}$ & 0.6 & 0.85 & 1.01 & 0.97 & 0.3 \EndAccSupp{} \\
        \BeginAccSupp{ActualText=0400-00700 30-50 01.4081 12.22 3.25 +4.47-3.68 03.51 0.83 -0.18 +6.28-0.05 0.59 0.982 1.01 0.97 0.5}    400--700 & 30--50 &  1&41  & 12.2 & 3.2 & ${}^{+4.5}_{-3.7}$ &  3.5 & 0.8 & -0.2 & ${}^{+6.3}_{-0.1}$ & 0.6 & 0.98 & 1.01 & 0.97 & 0.5 \EndAccSupp{} \\ \hline
        \BeginAccSupp{ActualText=0700-05000 07-11 42.8588 03.25 1.29 +2.46-2.23 02.39 1.07 +0.12 +0.39-0.17 0.51 0.826 1.03 0.90 0.6}   700--5000 &  7--11 & 42&9   &  3.3 & 1.3 & ${}^{+2.5}_{-2.2}$ &  2.4 & 1.1 &  0.1 & ${}^{+0.4}_{-0.2}$ & 0.5 & 0.83 & 1.03 & 0.90 & 0.6 \EndAccSupp{} \\
        \BeginAccSupp{ActualText=0700-05000 11-18 27.7541 02.80 1.83 +2.55-2.55 02.22 0.95 -0.03 +0.90-0.19 0.50 0.702 1.03 0.97 0.7}   700--5000 & 11--18 & 27&8   &  2.8 & 1.8 & ${}^{+2.5}_{-2.5}$ &  2.2 & 0.9 & -0.0 & ${}^{+0.9}_{-0.2}$ & 0.5 & 0.70 & 1.03 & 0.97 & 0.7 \EndAccSupp{} \\
        \BeginAccSupp{ActualText=0700-05000 18-30 12.0890 04.04 0.92 +2.66-2.71 03.52 0.92 -0.02 +1.19-0.01 0.50 0.571 1.03 0.97 0.0}   700--5000 & 18--30 & 12&1   &  4.0 & 0.9 & ${}^{+2.7}_{-2.7}$ &  3.5 & 0.9 & -0.0 & ${}^{+1.2}_{-0.0}$ & 0.5 & 0.57 & 1.03 & 0.97 & 0.0 \EndAccSupp{} \\
        \BeginAccSupp{ActualText=0700-05000 30-50 02.2274 09.38 1.55 +4.63-4.38 03.98 1.40 +0.07 -0.74-0.39 0.50 0.714 1.03 0.96 0.4}   700--5000 & 30--50 &  2&23  &  9.4 & 1.6 & ${}^{+4.6}_{-4.4}$ &  4.0 & 1.4 &  0.1 & ${}^{-0.7}_{-0.4}$ & 0.5 & 0.71 & 1.03 & 0.96 & 0.4 \EndAccSupp{} \\ \hline
        \BeginAccSupp{ActualText=5000-15000 07-11 02.4809 22.64 2.90 +1.57-1.58 -3.42 2.00 +0.14 -1.01-0.54 0.32 0.930 1.16 0.89 0.6} 5000--15000 &  7--11 &  2&48  & 22.6 & 2.9 & ${}^{+1.6}_{-1.6}$ & -3.4 & 2.0 &  0.1 & ${}^{-1.0}_{-0.5}$ & 0.3 & 0.93 & 1.16 & 0.89 & 0.6 \EndAccSupp{} \\
        \BeginAccSupp{ActualText=5000-15000 11-18 01.9849 13.59 3.47 +1.26-1.23 04.34 1.40 -0.05 -0.58+0.01 0.32 0.867 1.16 0.95 0.3} 5000--15000 & 11--18 &  1&99  & 13.6 & 3.5 & ${}^{+1.3}_{-1.2}$ &  4.3 & 1.4 & -0.1 & ${}^{-0.6}_{+0.0}$ & 0.3 & 0.87 & 1.16 & 0.95 & 0.3 \EndAccSupp{} \\
        \BeginAccSupp{ActualText=5000-15000 18-30 00.9652 14.78 3.10 +1.53-1.46 06.40 0.85 -0.03 -2.38-1.60 0.31 0.722 1.16 0.98 0.1} 5000--15000 & 18--30 &  0&965 & 14.8 & 3.1 & ${}^{+1.5}_{-1.5}$ &  6.4 & 0.8 & -0.0 & ${}^{-2.4}_{-1.6}$ & 0.3 & 0.72 & 1.16 & 0.98 & 0.1 \EndAccSupp{} \\
        \BeginAccSupp{ActualText=5000-15000 30-50 00.2038 32.51 4.89 +3.46-3.41 06.88 1.70 -0.08 -5.01-4.14 0.30 0.433 1.16 0.98 0.9} 5000--15000 & 30--50 &  0&204 & 32.5 & 4.9 & ${}^{+3.5}_{-3.4}$ &  6.9 & 1.7 & -0.1 & ${}^{-5.0}_{-4.1}$ & 0.3 & 0.43 & 1.16 & 0.98 & 0.9 \EndAccSupp{} \\ \hline
    \end{tabular}}
    \caption{Double-differential inclusive-jet cross sections, $\sigma$. Also listed are the unfolding uncertainty $\delta_\text{unf}$, the sum of the uncorrelated systematic uncertainties $\delta_\text{uncor}$ and the correlated systematic uncertainties associated with the jet-energy scale $\delta_\text{JES}$, the MC model $\delta_\text{model}$, the relative normalisation of the background from unmatched detector-level jets $\delta_\text{fake}$, the relative normalisation of the background from low-$Q^2$ DIS events $\delta_\text{Low-$Q^2$}$, the $(E-p_\text{Z})$-cut boundaries $\delta_{E-p_\text{Z}}$, the track-matching-efficiency correction $\delta_\text{TME}$. Uncertainties for which a single number is listed should be taken as symmetric in the other direction. Not listed explicitly is the luminosity uncertainty of $1.9\%$, which is fully correlated across all points. The last four columns show the QED Born-level correction $c_\text{QED}$ that has been applied to the data as well as the $Z$, $c_{Z}$, and hadronisation correction and associated uncertainty, $c_\text{Had}$ and $\delta_\text{Had}$, that need to be applied to the theory predictions.
    }
    \label{tab:cross sections}
    \vfill
\end{table}

\begin{table}[p]
    \vfill
    \centering
    \scalebox{0.8}{\begin{tabular}{|@{\,}>{\boldmath\bfseries}r>{\boldmath\bfseries}r|r@{\,}r@{\,}r@{\,}r@{\,}r@{\,}r@{\,}r@{\,}r@{\,}r@{\,}r@{\,}r@{\,}r@{\,}r@{\,}r|} \hline
        \multicolumn{1}{|c}{\!\boldmath\bfseries\makecell{$Q^2$\\$(\text{GeV}^2)$}\!} &
        \multicolumn{1}{c|}{\!\!\!\!\!\!\!\!\boldmath\bfseries\makecell{$p_{\perp,\text{Breit}}$\\$(\text{GeV})$}\!\!\!\!} &
        \multicolumn{1}{c}{\!\boldmath\bfseries\makecell{$\delta_\text{rew.}$\\$(\%)$}\!} &
        \multicolumn{1}{c}{\!\boldmath\bfseries\makecell{$\delta_\text{EES}$\\$(\%)$}\!} &
        \multicolumn{1}{c}{\!\boldmath\bfseries\makecell{$\delta_\text{EM}$\\$(\%)$}\!} &
        \multicolumn{1}{c}{\!\boldmath\bfseries\makecell{$\delta_\text{EL}$\\$(\%)$}\!} &
        \multicolumn{1}{c}{\!\boldmath\bfseries\makecell{$\delta_{p_T}$\\$(\%)$}\!} &
        \multicolumn{1}{c}{\!\boldmath\bfseries\makecell{$\delta_\text{trk.}$\\$(\%)$}\!} &
        \multicolumn{1}{c}{\!\boldmath\bfseries\makecell{$\delta_\text{bal.}$\\$(\%)$}\!} &
        \multicolumn{1}{c}{\!\boldmath\bfseries\makecell{$\delta_\text{vtx.}$\\$(\%)$}\!} &
        \multicolumn{1}{c}{\!\boldmath\bfseries\makecell{$\delta_\text{rad.}$\\$(\%)$}\!} &
        \multicolumn{1}{c}{\!\!\boldmath\bfseries\makecell{$\delta_\text{DCA}$\\$(\%)$}\!} &
        \multicolumn{1}{c}{\!\boldmath\bfseries\makecell{$\delta_\text{PHP}$\\$(\%)$}\!} &
        \multicolumn{1}{c}{\!\boldmath\bfseries\makecell{$\delta_\text{pol.}$\\$(\%)$}\!} &
        \multicolumn{1}{c}{\!\boldmath\bfseries\makecell{$\delta_\text{FLT}$\\$(\%)$}\!} &
        \multicolumn{1}{c|}{\!\boldmath\bfseries\makecell{$\delta_\text{QED}$\\$(\%)$}\!\!} \\ \hline
        \BeginAccSupp{ActualText=0150-00200 07-11 -0.88 -0.03+0.12 +0.59 -1.13 +0.42-0.57 +0.00+0.31 +0.00+0.04 +0.06-0.05 +0.00-0.00 -0.29 +0.01 -0.04 +0.04 +0.17}    150--200 &  7--11 & -0.9 & ${}^{-0.0}_{+0.1}$ & +0.6 & -1.1 & ${}^{+0.4}_{-0.6}$ & ${}^{+0.0}_{+0.3}$ & ${}^{+0.0}_{+0.0}$ & ${}^{+0.1}_{-0.0}$ & ${}^{+0.0}_{-0.0}$ & -0.3 & +0.0 & -0.0 & +0.0 & +0.2 \EndAccSupp{} \\
        \BeginAccSupp{ActualText=0150-00200 11-18 -1.22 -0.05+0.07 +0.24 -1.12 +0.48-0.01 +0.00+0.23 +0.00+0.02 -0.01+0.02 +0.01-0.02 -0.25 +0.07 -0.04 -0.02 +0.35}    150--200 & 11--18 & -1.2 & ${}^{-0.1}_{+0.1}$ & +0.2 & -1.1 & ${}^{+0.5}_{-0.0}$ & ${}^{+0.0}_{+0.2}$ & ${}^{+0.0}_{+0.0}$ & ${}^{-0.0}_{+0.0}$ & ${}^{+0.0}_{-0.0}$ & -0.3 & +0.1 & -0.0 & -0.0 & +0.3 \EndAccSupp{} \\
        \BeginAccSupp{ActualText=0150-00200 18-30 +0.48 -0.03+0.04 -0.38 -1.15 +0.16+0.09 +0.00-0.20 +0.07+0.16 -0.15-0.02 -0.06+0.03 -0.54 -0.13 -0.04 +0.01 +0.89}    150--200 & 18--30 & +0.5 & ${}^{-0.0}_{+0.0}$ & -0.4 & -1.1 & ${}^{+0.2}_{+0.1}$ & ${}^{+0.0}_{-0.2}$ & ${}^{+0.1}_{+0.2}$ & ${}^{-0.2}_{-0.0}$ & ${}^{-0.1}_{+0.0}$ & -0.5 & -0.1 & -0.0 & +0.0 & +0.9 \EndAccSupp{} \\
        \BeginAccSupp{ActualText=0150-00200 30-50 +1.43 +0.04+0.27 -3.03 +2.75 -0.06+0.21 +0.00+1.07 +0.42+0.14 -0.13-0.77 -0.11+0.00 -1.61 +0.03 -0.03 -0.13 +2.55}    150--200 & 30--50 & +1.4 & ${}^{+0.0}_{+0.3}$ & -3.0 & +2.8 & ${}^{-0.1}_{+0.2}$ & ${}^{+0.0}_{+1.1}$ & ${}^{+0.4}_{+0.1}$ & ${}^{-0.1}_{-0.8}$ & ${}^{-0.1}_{+0.0}$ & -1.6 & +0.0 & -0.0 & -0.1 & +2.5 \EndAccSupp{} \\ \hline
        \BeginAccSupp{ActualText=0200-00270 07-11 -0.68 -0.00+0.08 +0.76 -0.48 +1.02-1.08 +0.00+0.08 +0.03+0.04 -0.02+0.03 +0.11-0.01 -0.41 -0.06 -0.04 +0.04 +0.15}    200--270 &  7--11 & -0.7 & ${}^{-0.0}_{+0.1}$ & +0.8 & -0.5 & ${}^{+1.0}_{-1.1}$ & ${}^{+0.0}_{+0.1}$ & ${}^{+0.0}_{+0.0}$ & ${}^{-0.0}_{+0.0}$ & ${}^{+0.1}_{-0.0}$ & -0.4 & -0.1 & -0.0 & +0.0 & +0.1 \EndAccSupp{} \\
        \BeginAccSupp{ActualText=0200-00270 11-18 -1.18 -0.04+0.03 -0.21 -0.25 +0.13-0.23 +0.00-0.03 -0.02+0.03 -0.09+0.01 -0.07-0.09 -0.32 +0.15 -0.04 -0.04 +0.27}    200--270 & 11--18 & -1.2 & ${}^{-0.0}_{+0.0}$ & -0.2 & -0.3 & ${}^{+0.1}_{-0.2}$ & ${}^{+0.0}_{-0.0}$ & ${}^{-0.0}_{+0.0}$ & ${}^{-0.1}_{+0.0}$ & ${}^{-0.1}_{-0.1}$ & -0.3 & +0.1 & -0.0 & -0.0 & +0.3 \EndAccSupp{} \\
        \BeginAccSupp{ActualText=0200-00270 18-30 +0.19 -0.01-0.03 +2.57 +0.23 +0.02-0.21 +0.00+0.09 +0.01+0.13 -0.06-0.11 -0.20-0.84 -0.50 +0.07 -0.05 -0.01 +0.71}    200--270 & 18--30 & +0.2 & ${}^{-0.0}_{-0.0}$ & +2.6 & +0.2 & ${}^{+0.0}_{-0.2}$ & ${}^{+0.0}_{+0.1}$ & ${}^{+0.0}_{+0.1}$ & ${}^{-0.1}_{-0.1}$ & ${}^{-0.2}_{-0.8}$ & -0.5 & +0.1 & -0.0 & -0.0 & +0.7 \EndAccSupp{} \\
        \BeginAccSupp{ActualText=0200-00270 30-50 +1.90 -0.18+0.39 +5.99 +1.69 +0.00+0.23 +0.00+1.15 +0.00+0.74 -0.12+0.09 +0.11+1.21 -0.25 -0.24 -0.06 +0.03 +2.51}    200--270 & 30--50 & +1.9 & ${}^{-0.2}_{+0.4}$ & +6.0 & +1.7 & ${}^{+0.0}_{+0.2}$ & ${}^{+0.0}_{+1.1}$ & ${}^{+0.0}_{+0.7}$ & ${}^{-0.1}_{+0.1}$ & ${}^{+0.1}_{+1.2}$ & -0.3 & -0.2 & -0.1 & +0.0 & +2.5 \EndAccSupp{} \\ \hline
        \BeginAccSupp{ActualText=0270-00400 07-11 -0.66 -0.07+0.10 +1.20 -0.57 +0.59-0.73 +0.00+0.11 +0.03+0.03 -0.02-0.04 +0.09+0.14 -0.34 +0.21 -0.05 -0.10 +0.15}    270--400 &  7--11 & -0.7 & ${}^{-0.1}_{+0.1}$ & +1.2 & -0.6 & ${}^{+0.6}_{-0.7}$ & ${}^{+0.0}_{+0.1}$ & ${}^{+0.0}_{+0.0}$ & ${}^{-0.0}_{-0.0}$ & ${}^{+0.1}_{+0.1}$ & -0.3 & +0.2 & -0.1 & -0.1 & +0.1 \EndAccSupp{} \\
        \BeginAccSupp{ActualText=0270-00400 11-18 -1.50 -0.01+0.05 +1.47 -0.66 +0.38-0.56 +0.00+0.07 -0.01+0.04 +0.06+0.11 +0.05-0.11 -0.27 -0.00 -0.06 -0.18 +0.20}    270--400 & 11--18 & -1.5 & ${}^{-0.0}_{+0.0}$ & +1.5 & -0.7 & ${}^{+0.4}_{-0.6}$ & ${}^{+0.0}_{+0.1}$ & ${}^{-0.0}_{+0.0}$ & ${}^{+0.1}_{+0.1}$ & ${}^{+0.0}_{-0.1}$ & -0.3 & -0.0 & -0.1 & -0.2 & +0.2 \EndAccSupp{} \\
        \BeginAccSupp{ActualText=0270-00400 18-30 +0.38 +0.06+0.09 +0.58 -1.74 -0.14+0.14 +0.00+0.00 -0.01+0.25 +0.12-0.05 -0.52-0.22 -0.80 -0.05 -0.05 -0.31 +0.59}    270--400 & 18--30 & +0.4 & ${}^{+0.1}_{+0.1}$ & +0.6 & -1.7 & ${}^{-0.1}_{+0.1}$ & ${}^{+0.0}_{+0.0}$ & ${}^{-0.0}_{+0.2}$ & ${}^{+0.1}_{-0.0}$ & ${}^{-0.5}_{-0.2}$ & -0.8 & -0.1 & -0.1 & -0.3 & +0.6 \EndAccSupp{} \\
        \BeginAccSupp{ActualText=0270-00400 30-50 +0.67 -0.13+0.30 +3.64 -2.11 -0.09+0.00 +0.00-0.01 +0.21+0.32 -0.13-0.67 -1.01-1.91 +0.37 -0.00 -0.05 -0.91 +1.75}    270--400 & 30--50 & +0.7 & ${}^{-0.1}_{+0.3}$ & +3.6 & -2.1 & ${}^{-0.1}_{+0.0}$ & ${}^{+0.0}_{-0.0}$ & ${}^{+0.2}_{+0.3}$ & ${}^{-0.1}_{-0.7}$ & ${}^{-1.0}_{-1.9}$ & +0.4 & -0.0 & -0.1 & -0.9 & +1.8 \EndAccSupp{} \\ \hline
        \BeginAccSupp{ActualText=0400-00700 07-11 -0.67 -0.03+0.10 +0.20 -0.21 +0.55-0.73 +0.00-0.01 -0.00+0.03 +0.08-0.02 -0.17-0.12 -0.14 +0.21 -0.06 -0.29 +0.16}    400--700 &  7--11 & -0.7 & ${}^{-0.0}_{+0.1}$ & +0.2 & -0.2 & ${}^{+0.5}_{-0.7}$ & ${}^{+0.0}_{-0.0}$ & ${}^{-0.0}_{+0.0}$ & ${}^{+0.1}_{-0.0}$ & ${}^{-0.2}_{-0.1}$ & -0.1 & +0.2 & -0.1 & -0.3 & +0.2 \EndAccSupp{} \\
        \BeginAccSupp{ActualText=0400-00700 11-18 -1.06 +0.01+0.07 +0.13 -0.25 +0.24-0.15 +0.00+0.02 -0.05-0.03 -0.04+0.06 -0.16-0.07 -0.07 +0.07 -0.06 -0.27 +0.17}    400--700 & 11--18 & -1.1 & ${}^{+0.0}_{+0.1}$ & +0.1 & -0.3 & ${}^{+0.2}_{-0.2}$ & ${}^{+0.0}_{+0.0}$ & ${}^{-0.0}_{-0.0}$ & ${}^{-0.0}_{+0.1}$ & ${}^{-0.2}_{-0.1}$ & -0.1 & +0.1 & -0.1 & -0.3 & +0.2 \EndAccSupp{} \\
        \BeginAccSupp{ActualText=0400-00700 18-30 +0.53 +0.12+0.06 -0.37 +0.34 -0.02-0.06 +0.00+0.08 -0.09+0.27 +0.07-0.07 -0.52+0.24 -0.05 +0.30 -0.06 -0.19 +0.47}    400--700 & 18--30 & +0.5 & ${}^{+0.1}_{+0.1}$ & -0.4 & +0.3 & ${}^{-0.0}_{-0.1}$ & ${}^{+0.0}_{+0.1}$ & ${}^{-0.1}_{+0.3}$ & ${}^{+0.1}_{-0.1}$ & ${}^{-0.5}_{+0.2}$ & -0.0 & +0.3 & -0.1 & -0.2 & +0.5 \EndAccSupp{} \\
        \BeginAccSupp{ActualText=0400-00700 30-50 +1.45 +0.25-0.07 +1.40 +1.42 -0.21-0.72 +0.00-0.27 -0.16+0.37 -0.14+0.16 -0.45-1.60 +0.24 -0.00 -0.05 -0.88 +1.49}    400--700 & 30--50 & +1.5 & ${}^{+0.2}_{-0.1}$ & +1.4 & +1.4 & ${}^{-0.2}_{-0.7}$ & ${}^{+0.0}_{-0.3}$ & ${}^{-0.2}_{+0.4}$ & ${}^{-0.1}_{+0.2}$ & ${}^{-0.4}_{-1.6}$ & +0.2 & -0.0 & -0.0 & -0.9 & +1.5 \EndAccSupp{} \\ \hline
        \BeginAccSupp{ActualText=0700-05000 07-11 -0.23 -0.08+0.14 +0.98 -0.09 +0.97-0.47 +0.00+0.13 -0.01+0.02 +0.01-0.02 +0.00+0.00 -0.10 -0.06 -0.08 +0.29 +0.14}   700--5000 &  7--11 & -0.2 & ${}^{-0.1}_{+0.1}$ & +1.0 & -0.1 & ${}^{+1.0}_{-0.5}$ & ${}^{+0.0}_{+0.1}$ & ${}^{-0.0}_{+0.0}$ & ${}^{+0.0}_{-0.0}$ & ${}^{+0.0}_{+0.0}$ & -0.1 & -0.1 & -0.1 & +0.3 & +0.1 \EndAccSupp{} \\
        \BeginAccSupp{ActualText=0700-05000 11-18 -1.69 -0.05+0.09 +0.50 +0.20 +0.41-0.19 +0.00+0.15 -0.00+0.14 -0.02-0.04 +0.00+0.00 -0.13 +0.21 -0.09 +0.12 +0.14}   700--5000 & 11--18 & -1.7 & ${}^{-0.1}_{+0.1}$ & +0.5 & +0.2 & ${}^{+0.4}_{-0.2}$ & ${}^{+0.0}_{+0.2}$ & ${}^{-0.0}_{+0.1}$ & ${}^{-0.0}_{-0.0}$ & ${}^{+0.0}_{+0.0}$ & -0.1 & +0.2 & -0.1 & +0.1 & +0.1 \EndAccSupp{} \\
        \BeginAccSupp{ActualText=0700-05000 18-30 -0.77 -0.11+0.10 +0.07 +0.07 +0.23-0.50 +0.00+0.01 -0.08+0.10 -0.01-0.31 +0.00+0.00 -0.01 +0.17 -0.11 +0.03 +0.17}   700--5000 & 18--30 & -0.8 & ${}^{-0.1}_{+0.1}$ & +0.1 & +0.1 & ${}^{+0.2}_{-0.5}$ & ${}^{+0.0}_{+0.0}$ & ${}^{-0.1}_{+0.1}$ & ${}^{-0.0}_{-0.3}$ & ${}^{+0.0}_{+0.0}$ & -0.0 & +0.2 & -0.1 & +0.0 & +0.2 \EndAccSupp{} \\
        \BeginAccSupp{ActualText=0700-05000 30-50 -0.20 -0.33+0.46 +0.12 +0.69 +2.06+0.25 +0.00-0.28 +0.00+0.24 -0.07-0.58 +0.00+0.00 +0.04 -0.03 -0.09 -0.10 +0.49}   700--5000 & 30--50 & -0.2 & ${}^{-0.3}_{+0.5}$ & +0.1 & +0.7 & ${}^{+2.1}_{+0.2}$ & ${}^{+0.0}_{-0.3}$ & ${}^{+0.0}_{+0.2}$ & ${}^{-0.1}_{-0.6}$ & ${}^{+0.0}_{+0.0}$ & +0.0 & -0.0 & -0.1 & -0.1 & +0.5 \EndAccSupp{} \\ \hline
        \BeginAccSupp{ActualText=5000-15000 07-11 -0.03 -0.15+0.15 +1.49 -1.55 +0.08-2.25 +0.00+0.05 +0.02-0.31 +0.11-0.24 +0.00+0.00 -0.11 +0.30 -0.07 +1.49 +0.19} 5000--15000 &  7--11 & -0.0 & ${}^{-0.2}_{+0.2}$ & +1.5 & -1.6 & ${}^{+0.1}_{-2.2}$ & ${}^{+0.0}_{+0.1}$ & ${}^{+0.0}_{-0.3}$ & ${}^{+0.1}_{-0.2}$ & ${}^{+0.0}_{+0.0}$ & -0.1 & +0.3 & -0.1 & +1.5 & +0.2 \EndAccSupp{} \\
        \BeginAccSupp{ActualText=5000-15000 11-18 -1.20 -0.17+0.16 -2.69 +0.80 +0.19-1.11 +0.00-0.05 +0.21-0.95 -0.19-0.47 +0.00+0.00 -0.78 -0.17 -0.10 +1.06 +0.18} 5000--15000 & 11--18 & -1.2 & ${}^{-0.2}_{+0.2}$ & -2.7 & +0.8 & ${}^{+0.2}_{-1.1}$ & ${}^{+0.0}_{-0.0}$ & ${}^{+0.2}_{-1.0}$ & ${}^{-0.2}_{-0.5}$ & ${}^{+0.0}_{+0.0}$ & -0.8 & -0.2 & -0.1 & +1.1 & +0.2 \EndAccSupp{} \\
        \BeginAccSupp{ActualText=5000-15000 18-30 -1.83 -0.16+0.15 +2.04 -0.07 +1.07+1.11 +0.00-0.04 -0.06-0.45 -0.19-0.69 +0.00+0.00 -0.25 -0.11 -0.07 +0.70 +0.21} 5000--15000 & 18--30 & -1.8 & ${}^{-0.2}_{+0.1}$ & +2.0 & -0.1 & ${}^{+1.1}_{+1.1}$ & ${}^{+0.0}_{-0.0}$ & ${}^{-0.1}_{-0.4}$ & ${}^{-0.2}_{-0.7}$ & ${}^{+0.0}_{+0.0}$ & -0.3 & -0.1 & -0.1 & +0.7 & +0.2 \EndAccSupp{} \\
        \BeginAccSupp{ActualText=5000-15000 30-50 -1.53 -0.19+0.23 +4.52 +0.13 +1.01-0.25 +0.00-0.06 -0.07+0.14 -0.17+0.23 +0.00+0.00 +0.37 +0.63 +0.20 +0.11 +0.24} 5000--15000 & 30--50 & -1.5 & ${}^{-0.2}_{+0.2}$ & +4.5 & +0.1 & ${}^{+1.0}_{-0.3}$ & ${}^{+0.0}_{-0.1}$ & ${}^{-0.1}_{+0.1}$ & ${}^{-0.2}_{+0.2}$ & ${}^{+0.0}_{+0.0}$ & +0.4 & +0.6 & +0.2 & +0.1 & +0.2 \EndAccSupp{} \\ \hline
    \end{tabular}}
    \caption{Breakdown of the uncorrelated uncertainty $\delta_\text{uncor}$ from Table~\ref{tab:cross sections}. Shown are the uncertainties associated with the reweighting of the MC models ($\delta_\text{rew.}$), the electron-energy scale ($\delta_\text{EES}$), the electron-finding algorithm ($\delta_\text{EM}$), the electron calibration ($\delta_\text{EL}$), the variation of the $p_{T,\text{lab}}$ cut of the jets ($\delta_{p_T}$), the variation of the electron-track momentum-cut boundaries ($\delta_\text{trk.}$), the variation of the $p_T/\sqrt{E_T}$-cut boundaries ($\delta_\text{bal.}$), the variation of the $Z_\text{vertex}$-cut boundaries ($\delta_\text{vtx.}$), the variation of the $R_\text{RCAL}$-cut boundaries ($\delta_\text{rad.}$), the variation of the electron-track distance-cut boundaries ($\delta_\text{DCA}$), the relative normalisation of the background from photoproduction events ($\delta_\text{PHP}$), the polarisation correction ($\delta_\text{pol.}$), the FLT track-veto-efficiency correction ($\delta_\text{FLT}$) and the correction to QED Born-level ($\delta_\text{QED}$). For the asymmetric uncertainties, the upper number corresponds to the upward variation of the corresponding parameter and the lower number corresponds to the downward variation.
    }
    \label{tab:uncertainties}
    \vfill
\end{table}

\begin{table}[p]
    \vfill
    \centering
    \newcommand{\rcell}[1]{\multirow[c]{1}[24]{*}{\rotatebox{90}{\parbox{2.65cm}{\hfill\bfseries #1~}}}}
    \def\arraystretch{1.125}
    \scalebox{0.635}{\begin{tabular}{|>{\bfseries\boldmath}c|>{\bfseries\boldmath}r>{\bfseries\boldmath}r|%
        >{\!$}r<{$\!}@{~}>{\!$}r<{$\!}@{~}>{\!$}r<{$\!}@{~}>{\!$}r<{$\!}|%
        >{\!$}r<{$\!}@{~}>{\!$}r<{$\!}@{~}>{\!$}r<{$\!}@{~}>{\!$}r<{$\!}|%
        >{\!$}r<{$\!}@{~}>{\!$}r<{$\!}@{~}>{\!$}r<{$\!}@{~}>{\!$}r<{$\!}|%
        >{\!$}r<{$\!}@{~}>{\!$}r<{$\!}@{~}>{\!$}r<{$\!}@{~}>{\!$}r<{$\!}|%
        >{\!$}r<{$\!}@{~}>{\!$}r<{$\!}@{~}>{\!$}r<{$\!}@{~}>{\!$}r<{$\!}|%
        >{\!$}r<{$\!}@{~}>{\!$}r<{$\!}@{~}>{\!$}r<{$\!}@{~}>{\!$}r<{$\!}|} \cline{4-27}
        \multicolumn{1}{c}{} &&& \multicolumn{24}{c|}{\bfseries Inclusive jet bin} \\ \cline{4-27}
        \multicolumn{1}{c}{} & \multirow[t]{1}[20]{*}{\makecell{\\[1cm] $Q^2$ \\ (GeV$^2$)}} &&
        \rcell{150--200~} &    \rcell{150--200~} &    \rcell{150--200~} &    \rcell{150--200~} &
        \rcell{200--270~} &    \rcell{200--270~} &    \rcell{200--270~} &    \rcell{200--270~} &
        \rcell{270--400~} &    \rcell{270--400~} &    \rcell{270--400~} &    \rcell{270--400~} &
        \rcell{400--700~} &    \rcell{400--700~} &    \rcell{400--700~} &    \rcell{400--700~} &
        \rcell{700--5000~} &   \rcell{700--5000~} &   \rcell{700--5000~} &   \rcell{700--5000~} &
        \rcell{5000--15000~} & \rcell{5000--15000~} & \rcell{5000--15000~} & \rcell{5000--15000~} \\[2.325cm]
        \multicolumn{1}{c}{} && \multirow[t]{1}[20]{*}{\makecell{\\[0.125cm] $p_\perp$ \\ (GeV)}} &
        \rcell{7--11} & \rcell{11--18} & \rcell{18--30} & \rcell{30--50} &
        \rcell{7--11} & \rcell{11--18} & \rcell{18--30} & \rcell{30--50} &
        \rcell{7--11} & \rcell{11--18} & \rcell{18--30} & \rcell{30--50} &
        \rcell{7--11} & \rcell{11--18} & \rcell{18--30} & \rcell{30--50} &
        \rcell{7--11} & \rcell{11--18} & \rcell{18--30} & \rcell{30--50} &
        \rcell{7--11} & \rcell{11--18} & \rcell{18--30} & \rcell{30--50} \\[0.875cm] \hline
        \multirow{24}{*}{\BeginAccSupp{ActualText=}\rotatebox{90}{Inclusive jet bin}\EndAccSupp{}}
        \BeginAccSupp{ActualText=0150-00200 07-11 100 -38 -07 +00 -13 +06 +02 +00 +13 -05 -01 -00 +09 -03 -01 -00 +09 -04 -02 -01 +02 -01 -01 -01} &    150--200 &  7--11 &100&-38& -7&  0&-13&  6&  2&  0& 13& -5& -1& -0&  9& -3& -1& -0&  9& -4& -2& -1&  2& -1& -1& -1 \EndAccSupp{} \\
        \BeginAccSupp{ActualText=0150-00200 11-18 -38 100 +04 +04 +05 -26 -00 -01 -04 +07 +01 +01 -03 +01 +00 -00 -03 +02 +01 +00 -01 +00 +00 +00} &    150--200 & 11--18 &-38&100&  4&  4&  5&-26& -0& -1& -4&  7&  1&  1& -3&  1&  0& -0& -3&  2&  1&  0& -1&  0&  0&  0 \EndAccSupp{} \\
        \BeginAccSupp{ActualText=0150-00200 18-30 -07 +04 100 -01 +00 +03 -36 +01 -00 -00 +09 -00 -00 +00 -01 +00 -00 +00 +00 +00 -00 +00 +00 +00} &    150--200 & 18--30 & -7&  4&100& -1&  0&  3&-36&  1& -0& -0&  9& -0& -0&  0& -1&  0& -0&  0&  0&  0& -0&  0&  0&  0 \EndAccSupp{} \\
        \BeginAccSupp{ActualText=0150-00200 30-50 +00 +04 -01 100 +00 -00 +02 -47 +00 +00 -01 +14 +00 -00 +00 -03 +00 +00 +00 +00 +00 -00 -00 -00} &    150--200 & 30--50 &  0&  4& -1&100&  0& -0&  2&-47&  0&  0& -1& 14&  0& -0&  0& -3&  0&  0&  0&  0&  0& -0& -0& -0 \EndAccSupp{} \\ \cline{2-27}
        \BeginAccSupp{ActualText=0200-00270 07-11 -13 +05 +00 +00 100 -41 -08 -01 -10 +04 +02 +00 +09 -03 -01 -00 +07 -03 -01 -01 +02 -01 -01 -00} &    200--270 &  7--11 &-13&  5&  0&  0&100&-41& -8& -1&-10&  4&  2&  0&  9& -3& -1& -0&  7& -3& -1& -1&  2& -1& -1& -0 \EndAccSupp{} \\
        \BeginAccSupp{ActualText=0200-00270 11-18 +06 -26 +03 -00 -41 100 -04 -01 +03 -17 +01 +00 -03 +04 +00 -00 -03 +01 +00 +00 -01 +00 +00 +00} &    200--270 & 11--18 &  6&-26&  3& -0&-41&100& -4& -1&  3&-17&  1&  0& -3&  4&  0& -0& -3&  1&  0&  0& -1&  0&  0&  0 \EndAccSupp{} \\
        \BeginAccSupp{ActualText=0200-00270 18-30 +02 -00 -36 +02 -08 -04 100 -05 -01 +02 -26 +02 -01 -00 +05 -00 -01 +00 -00 +00 -00 +00 +00 +00} &    200--270 & 18--30 &  2& -0&-36&  2& -8& -4&100& -5& -1&  2&-26&  2& -1& -0&  5& -0& -1&  0& -0&  0& -0&  0&  0&  0 \EndAccSupp{} \\
        \BeginAccSupp{ActualText=0200-00270 30-50 +00 -01 +01 -47 -01 -01 -05 100 -00 +00 +03 -39 -00 +00 -01 +10 -00 +00 +00 -01 -00 +00 +00 +00} &    200--270 & 30--50 &  0& -1&  1&-47& -1& -1& -5&100& -0&  0&  3&-39& -0&  0& -1& 10& -0&  0&  0& -1& -0&  0&  0&  0 \EndAccSupp{} \\ \cline{2-27}
        \BeginAccSupp{ActualText=0270-00400 07-11 +13 -04 -00 +00 -10 +03 -01 -00 100 -38 -08 -01 -03 +01 +01 +00 +07 -03 -01 -01 +02 -01 -00 -00} &    270--400 &  7--11 & 13& -4& -0&  0&-10&  3& -1& -0&100&-38& -8& -1& -3&  1&  1&  0&  7& -3& -1& -1&  2& -1& -0& -0 \EndAccSupp{} \\
        \BeginAccSupp{ActualText=0270-00400 11-18 -05 +07 -00 +00 +04 -17 +02 +00 -38 100 -04 -01 +00 -09 +01 +00 -02 +02 +00 +00 -01 +00 +00 +00} &    270--400 & 11--18 & -5&  7& -0&  0&  4&-17&  2&  0&-38&100& -4& -1&  0& -9&  1&  0& -2&  2&  0&  0& -1&  0&  0&  0 \EndAccSupp{} \\
        \BeginAccSupp{ActualText=0270-00400 18-30 -01 +01 +09 -01 +02 +01 -26 +03 -08 -04 100 -03 -00 +02 -16 +01 -00 +00 +01 -00 -00 +00 +00 +00} &    270--400 & 18--30 & -1&  1&  9& -1&  2&  1&-26&  3& -8& -4&100& -3& -0&  2&-16&  1& -0&  0&  1& -0& -0&  0&  0&  0 \EndAccSupp{} \\
        \BeginAccSupp{ActualText=0270-00400 30-50 -00 +01 -00 +14 +00 +00 +02 -39 -01 -01 -03 100 -00 +00 +02 -25 -00 +00 -00 +03 -00 +00 -00 -00} &    270--400 & 30--50 & -0&  1& -0& 14&  0&  0&  2&-39& -1& -1& -3&100& -0&  0&  2&-25& -0&  0& -0&  3& -0&  0& -0& -0 \EndAccSupp{} \\ \cline{2-27}
        \BeginAccSupp{ActualText=0400-00700 07-11 +09 -03 -00 +00 +09 -03 -01 -00 -03 +00 -00 -00 100 -39 -09 -01 +01 -00 -00 -00 +02 -01 -00 -00} &    400--700 &  7--11 &  9& -3& -0&  0&  9& -3& -1& -0& -3&  0& -0& -0&100&-39& -9& -1&  1& -0& -0& -0&  2& -1& -0& -0 \EndAccSupp{} \\
        \BeginAccSupp{ActualText=0400-00700 11-18 -03 +01 +00 -00 -03 +04 -00 +00 +01 -09 +02 +00 -39 100 -05 -01 -01 -04 +01 +00 -01 +00 +00 +00} &    400--700 & 11--18 & -3&  1&  0& -0& -3&  4& -0&  0&  1& -9&  2&  0&-39&100& -5& -1& -1& -4&  1&  0& -1&  0&  0&  0 \EndAccSupp{} \\
        \BeginAccSupp{ActualText=0400-00700 18-30 -01 +00 -01 +00 -01 +00 +05 -01 +01 +01 -16 +02 -09 -05 100 -05 -00 +01 -07 +01 -00 +00 +00 +00} &    400--700 & 18--30 & -1&  0& -1&  0& -1&  0&  5& -1&  1&  1&-16&  2& -9& -5&100& -5& -0&  1& -7&  1& -0&  0&  0&  0 \EndAccSupp{} \\
        \BeginAccSupp{ActualText=0400-00700 30-50 -00 -00 +00 -03 -00 -00 -00 +10 +00 +00 +01 -25 -01 -01 -05 100 -00 +00 +01 -10 -00 +00 +00 +00} &    400--700 & 30--50 & -0& -0&  0& -3& -0& -0& -0& 10&  0&  0&  1&-25& -1& -1& -5&100& -0&  0&  1&-10& -0&  0&  0&  0 \EndAccSupp{} \\ \cline{2-27}
        \BeginAccSupp{ActualText=0700-05000 07-11 +09 -03 -00 +00 +07 -03 -01 -00 +07 -02 -00 -00 +01 -01 -00 -00 100 -50 -10 -02 +01 -00 -00 -00} &   700--5000 &  7--11 &  9& -3& -0&  0&  7& -3& -1& -0&  7& -2& -0& -0&  1& -1& -0& -0&100&-50&-10& -2&  1& -0& -0& -0 \EndAccSupp{} \\
        \BeginAccSupp{ActualText=0700-05000 11-18 -04 +02 +00 +00 -03 +01 +00 +00 -03 +02 +00 +00 -00 -04 +01 +00 -50 100 -10 -02 -00 -01 +00 +00} &   700--5000 & 11--18 & -4&  2&  0&  0& -3&  1&  0&  0& -3&  2&  0&  0& -0& -4&  1&  0&-50&100&-10& -2& -0& -1&  0&  0 \EndAccSupp{} \\
        \BeginAccSupp{ActualText=0700-05000 18-30 -02 +01 +00 +00 -01 +00 -00 +00 -01 +00 +01 -00 -00 +01 -07 +01 -10 -10 100 -07 -00 +00 -01 -00} &   700--5000 & 18--30 & -2&  1&  0&  0& -1&  0& -0&  0& -1&  0&  1& -0& -0&  1& -7&  1&-10&-10&100& -7& -0&  0& -1& -0 \EndAccSupp{} \\
        \BeginAccSupp{ActualText=0700-05000 30-50 -01 +00 +00 +00 -01 +00 +00 -01 -01 +00 -00 +03 -00 +00 +01 -10 -02 -02 -07 100 -00 -00 -00 -01} &   700--5000 & 30--50 & -1&  0&  0&  0& -1&  0&  0& -1& -1&  0& -0&  3& -0&  0&  1&-10& -2& -2& -7&100& -0& -0& -0& -1 \EndAccSupp{} \\ \cline{2-27}
        \BeginAccSupp{ActualText=5000-15000 07-11 +02 -01 -00 +00 +02 -01 -00 -00 +02 -01 -00 -00 +02 -01 -00 -00 +01 -00 -00 -00 100 -67 -15 -03} & 5000--15000 &  7--11 &  2& -1& -0&  0&  2& -1& -0& -0&  2& -1& -0& -0&  2& -1& -0& -0&  1& -0& -0& -0&100&-67&-15& -3 \EndAccSupp{} \\
        \BeginAccSupp{ActualText=5000-15000 11-18 -01 +00 +00 -00 -01 +00 +00 +00 -01 +00 +00 +00 -01 +00 +00 +00 -00 -01 +00 -00 -67 100 -12 +01} & 5000--15000 & 11--18 & -1&  0&  0& -0& -1&  0&  0&  0& -1&  0&  0&  0& -1&  0&  0&  0& -0& -1&  0& -0&-67&100&-12&  1 \EndAccSupp{} \\
        \BeginAccSupp{ActualText=5000-15000 18-30 -01 +00 +00 -00 -01 +00 +00 +00 -00 +00 +00 -00 -00 +00 +00 +00 -00 +00 -01 -00 -15 -12 100 -17} & 5000--15000 & 18--30 & -1&  0&  0& -0& -1&  0&  0&  0& -0&  0&  0& -0& -0&  0&  0&  0& -0&  0& -1& -0&-15&-12&100&-17 \EndAccSupp{} \\
        \BeginAccSupp{ActualText=5000-15000 30-50 -01 +00 +00 -00 -00 +00 +00 +00 -00 +00 +00 -00 -00 +00 +00 +00 -00 +00 -00 -01 -03 +01 -17 100} & 5000--15000 & 30--50 & -1&  0&  0& -0& -0&  0&  0&  0& -0&  0&  0& -0& -0&  0&  0&  0& -0&  0& -0& -1& -3&  1&-17&100 \EndAccSupp{} \\ \hline \hline
        \multirow{22}{*}{\BeginAccSupp{ActualText=}\rotatebox{90}{Dijet bin}\EndAccSupp{}}
        \BeginAccSupp{ActualText=0125-00250 08-15 +02 +26 +05 +03 -01 +11 -00 -01 -07 +02 +01 +01 -06 +03 +01 -00 -06 +03 +01 +01 -02 +01 +00 +00} &    125--250 &  8--15 &  2& 26&  5&  3& -1& 11& -0& -1& -7&  2&  1&  1& -6&  3&  1& -0& -6&  3&  1&  1& -2&  1&  0&  0 \EndAccSupp{} \\
        \BeginAccSupp{ActualText=0125-00250 15-22 +02 +13 +18 +02 +01 +04 +07 -01 -00 -01 -01 +00 -00 +00 +00 -00 -00 +00 +00 +00 -00 +00 +00 +00} &    125--250 & 15--22 &  2& 13& 18&  2&  1&  4&  7& -1& -0& -1& -1&  0& -0&  0&  0& -0& -0&  0&  0&  0& -0&  0&  0&  0 \EndAccSupp{} \\
        \BeginAccSupp{ActualText=0125-00250 22-30 -00 +04 +20 +05 -01 -00 +10 -01 -00 +00 -02 +00 -00 +00 +00 -00 -00 +00 +00 +00 -00 +00 +00 +00} &    125--250 & 22--30 & -0&  4& 20&  5& -1& -0& 10& -1& -0&  0& -2&  0& -0&  0&  0& -0& -0&  0&  0&  0& -0&  0&  0&  0 \EndAccSupp{} \\
        \BeginAccSupp{ActualText=0125-00250 30-60 +01 +04 +02 +36 -00 -00 -01 +10 -00 +00 +00 -02 -00 +00 +00 +00 -00 +00 +00 +00 -00 +00 +00 +00} &    125--250 & 30--60 &  1&  4&  2& 36& -0& -0& -1& 10& -0&  0&  0& -2& -0&  0&  0&  0& -0&  0&  0&  0& -0&  0&  0&  0 \EndAccSupp{} \\ \cline{2-27}
        \BeginAccSupp{ActualText=0250-00500 08-15 -09 +02 +01 +00 -05 +06 +01 +00 +04 +19 +00 +00 -04 +07 +00 +00 -07 +03 +01 +01 -02 +01 +00 +00} &    250--500 &  8--15 & -9&  2&  1&  0& -5&  6&  1&  0&  4& 19&  0&  0& -4&  7&  0&  0& -7&  3&  1&  1& -2&  1&  0&  0 \EndAccSupp{} \\
        \BeginAccSupp{ActualText=0250-00500 15-22 -01 +00 -01 +00 -01 +02 +02 +00 +01 +09 +15 +00 -01 +02 +05 +00 -01 +00 -00 +00 -00 +00 +00 +00} &    250--500 & 15--22 & -1&  0& -1&  0& -1&  2&  2&  0&  1&  9& 15&  0& -1&  2&  5&  0& -1&  0& -0&  0& -0&  0&  0&  0 \EndAccSupp{} \\
        \BeginAccSupp{ActualText=0250-00500 22-30 -01 +01 -01 +00 -01 +00 +03 -00 -01 +00 +21 +00 -01 +00 +06 +01 -01 +00 -01 +00 -00 +00 +00 +00} &    250--500 & 22--30 & -1&  1& -1&  0& -1&  0&  3& -0& -1&  0& 21&  0& -1&  0&  6&  1& -1&  0& -1&  0& -0&  0&  0&  0 \EndAccSupp{} \\
        \BeginAccSupp{ActualText=0250-00500 30-60 -00 +01 +00 -01 -00 +00 -00 +02 -01 +00 +01 +33 -00 +00 +00 +10 -00 +00 +00 -01 -00 +00 +00 +00} &    250--500 & 30--60 & -0&  1&  0& -1& -0&  0& -0&  2& -1&  0&  1& 33& -0&  0&  0& 10& -0&  0&  0& -1& -0&  0&  0&  0 \EndAccSupp{} \\ \cline{2-27}
        \BeginAccSupp{ActualText=0500-01000 08-15 -08 +03 +00 -00 -06 +03 +01 +00 -07 -01 +01 +00 +04 +19 -00 +00 +01 +13 +01 +01 -02 +01 +00 +00} &   500--1000 &  8--15 & -8&  3&  0& -0& -6&  3&  1&  0& -7& -1&  1&  0&  4& 19& -0&  0&  1& 13&  1&  1& -2&  1&  0&  0 \EndAccSupp{} \\
        \BeginAccSupp{ActualText=0500-01000 15-22 -02 +01 -00 +00 -01 +01 +01 -00 -02 -01 -04 +00 +01 +10 +18 +01 -01 +06 +09 -00 -00 +00 -00 -00} &   500--1000 & 15--22 & -2&  1& -0&  0& -1&  1&  1& -0& -2& -1& -4&  0&  1& 10& 18&  1& -1&  6&  9& -0& -0&  0& -0& -0 \EndAccSupp{} \\
        \BeginAccSupp{ActualText=0500-01000 22-30 -01 +01 -00 -00 -01 +00 +01 +00 -01 +00 -04 -01 -01 -00 +22 +02 -01 -00 +14 -00 -00 +00 -00 +00} &   500--1000 & 22--30 & -1&  1& -0& -0& -1&  0&  1&  0& -1&  0& -4& -1& -1& -0& 22&  2& -1& -0& 14& -0& -0&  0& -0&  0 \EndAccSupp{} \\
        \BeginAccSupp{ActualText=0500-01000 30-60 -01 +00 +00 -01 -01 +00 -00 +04 -01 +00 +00 -10 -01 +00 -01 +37 -01 +00 -01 +22 -00 +00 -00 -00} &   500--1000 & 30--60 & -1&  0&  0& -1& -1&  0& -0&  4& -1&  0&  0&-10& -1&  0& -1& 37& -1&  0& -1& 22& -0&  0& -0& -0 \EndAccSupp{} \\ \cline{2-27}
        \BeginAccSupp{ActualText=1000-02000 08-15 -06 +02 +00 -00 -05 +02 +00 +00 -04 +02 +00 +00 -05 -00 +01 +00 +06 +24 -01 +00 -01 +00 +00 +00} &  1000--2000 &  8--15 & -6&  2&  0& -0& -5&  2&  0&  0& -4&  2&  0&  0& -5& -0&  1&  0&  6& 24& -1&  0& -1&  0&  0&  0 \EndAccSupp{} \\
        \BeginAccSupp{ActualText=1000-02000 15-22 -02 +01 +00 +00 -01 +00 -00 +00 -01 +01 +01 -00 -02 -00 -02 +00 +01 +13 +17 +00 -00 -00 -00 -00} &  1000--2000 & 15--22 & -2&  1&  0&  0& -1&  0& -0&  0& -1&  1&  1& -0& -2& -0& -2&  0&  1& 13& 17&  0& -0& -0& -0& -0 \EndAccSupp{} \\
        \BeginAccSupp{ActualText=1000-02000 22-30 -01 +00 +00 +00 -01 +00 -00 -00 -01 +00 +00 +00 -01 +00 -02 -01 -01 -00 +25 +01 -00 +00 -01 -00} &  1000--2000 & 22--30 & -1&  0&  0&  0& -1&  0& -0& -0& -1&  0&  0&  0& -1&  0& -2& -1& -1& -0& 25&  1& -0&  0& -1& -0 \EndAccSupp{} \\
        \BeginAccSupp{ActualText=1000-02000 30-60 -01 +01 +00 +00 -01 +00 +00 -01 -01 +00 +00 +01 -01 +00 +00 -05 -00 +01 -01 +42 -00 +00 -00 -01} &  1000--2000 & 30--60 & -1&  1&  0&  0& -1&  0&  0& -1& -1&  0&  0&  1& -1&  0&  0& -5& -0&  1& -1& 42& -0&  0& -0& -1 \EndAccSupp{} \\ \cline{2-27}
        \BeginAccSupp{ActualText=2000-05000 08-16 -04 +01 +00 -00 -03 +01 +00 +00 -03 +01 +00 -00 -04 -00 +00 +00 +03 +20 -01 -00 -01 +01 +00 +00} &  2000--5000 &  8--16 & -4&  1&  0& -0& -3&  1&  0&  0& -3&  1&  0& -0& -4& -0&  0&  0&  3& 20& -1& -0& -1&  1&  0&  0 \EndAccSupp{} \\
        \BeginAccSupp{ActualText=2000-05000 16-28 -02 +01 +00 +00 -01 +00 -00 -00 -01 +00 +01 +00 -01 +00 -03 -00 +00 +04 +28 +00 -00 +00 +00 -00} &  2000--5000 & 16--28 & -2&  1&  0&  0& -1&  0& -0& -0& -1&  0&  1&  0& -1&  0& -3& -0&  0&  4& 28&  0& -0&  0&  0& -0 \EndAccSupp{} \\
        \BeginAccSupp{ActualText=2000-05000 28-60 -01 +00 +00 +00 -01 +00 +00 -00 -01 +00 +00 +01 -01 +00 -00 -03 -01 +01 +03 +25 -00 +00 -00 +00} &  2000--5000 & 28--60 & -1&  0&  0&  0& -1&  0&  0& -0& -1&  0&  0&  1& -1&  0& -0& -3& -1&  1&  3& 25& -0&  0& -0&  0 \EndAccSupp{} \\ \cline{2-27}
        \BeginAccSupp{ActualText=5000-20000 08-16 -02 +01 +00 -00 -02 +01 +00 +00 -02 +01 +00 +00 -02 +01 +00 +00 -02 +01 +00 +00 +08 +24 -03 +00} & 5000--20000 &  8--16 & -2&  1&  0& -0& -2&  1&  0&  0& -2&  1&  0&  0& -2&  1&  0&  0& -2&  1&  0&  0&  8& 24& -3&  0 \EndAccSupp{} \\
        \BeginAccSupp{ActualText=5000-20000 16-28 -01 +00 +00 -00 -01 +00 +00 +00 -01 +00 +00 +00 -01 +00 +00 +00 -01 +00 -00 +00 +03 +09 +31 -01} & 5000--20000 & 16--28 & -1&  0&  0& -0& -1&  0&  0&  0& -1&  0&  0&  0& -1&  0&  0&  0& -1&  0& -0&  0&  3&  9& 31& -1 \EndAccSupp{} \\
        \BeginAccSupp{ActualText=5000-20000 28-60 -01 +00 +00 -00 -01 +00 +00 +00 -01 +00 +00 -00 -01 +00 +00 +00 -01 +00 +00 -00 +00 +01 +07 +39} & 5000--20000 & 28--60 & -1&  0&  0& -0& -1&  0&  0&  0& -1&  0&  0& -0& -1&  0&  0&  0& -1&  0&  0& -0&  0&  1&  7& 39 \EndAccSupp{} \\ \hline
    \end{tabular}}
    \caption{Correlation matrix of the unfolding uncertainty within the inclusive-jet cross-section measurement and between the inclusive-jet measurement and the previous dijet measurement\protect\cite{zeusdijets}. Correlations are given in percent. The transverse momentum $p_\perp$ is $p_{\perp,\text{Breit}}$ for the inclusive jets and $\overline{p_{\perp,\text{Breit}}}$ for the dijets.}
    \label{tab:correlations}
    \vfill
\end{table}

\begin{table}[p]
    \vfill
    \begin{center}
        \def\arraystretch{1.25}
        \begin{tabular}{|l|>{\raggedleft}p{2.5cm}@{/}>{\raggedright\arraybackslash}p{2.5cm}|} \hline
            \multicolumn{1}{|c|}{\textbf{Dataset}} & \textbf{Partial \bm{$\chi^2$}} & \textbf{Number of points} \\ \hline
            HERA NC $e^+p$ DIS, $E_p = \unit[920]{\GeV}$ & 448 & 377 \\
            HERA NC $e^+p$ DIS, $E_p = \unit[820]{\GeV}$ &  65 &  70 \\
            HERA NC $e^+p$ DIS, $E_p = \unit[575]{\GeV}$ & 219 & 254 \\
            HERA NC $e^+p$ DIS, $E_p = \unit[460]{\GeV}$ & 217 & 204 \\
            HERA NC $e^-p$ DIS, $E_p = \unit[920]{\GeV}$ & 220 & 159 \\
            HERA CC $e^+p$ DIS, $E_p = \unit[920]{\GeV}$ &  48 &  39 \\
            HERA CC $e^-p$ DIS, $E_p = \unit[920]{\GeV}$ &  52 &  42 \\
            ZEUS HERA~I inclusive jets                   &  26 &  30 \\
            ZEUS HERA~I/II dijets                        &  15 &  16 \\
            ZEUS HERA~II inclusive jets                  &  15 &  24 \\ \hline
            Correlated $\chi^2$                          & \multicolumn{2}{c|}{96} \\
            Global $\chi^2$ per degree of freedom        & \multicolumn{2}{c|}{1419/1200} \\ \hline
        \end{tabular}
    \end{center}
    \caption{The partial $\chi^2$ values from the nominal fit at NNLO and the number of data points for each dataset.}
    \label{tab:chi2}
    \vfill
\end{table}

\begin{table}[p]
    \vfill
    \begin{center}
        \def\arraystretch{1.25}
        \begin{tabular}{|cc|c@{~}l@{~}l@{~}l|r@{~}l|} \hline
            \boldmath\bfseries\makecell{Number of \\ jet cross \\ sections} &
            \boldmath\bfseries\makecell{$\langle\mu\rangle$\\$(\unit{\GeV})$} &
            \boldmath\bfseries\makecell{$\alpha_s(M_Z^2)$} &
            \boldmath\bfseries\makecell{$\pm \delta_\text{exp./fit}$} &
            \boldmath\bfseries\makecell{$\pm \delta_\text{mod./par.}$} &
            \boldmath\bfseries\makecell{\,$\pm \delta_\text{scale}$} &
            \boldmath\bfseries\makecell{$\alpha_s(\langle\mu\rangle^2)$\!} &
            \boldmath\bfseries\makecell{\!\!$\pm \delta_\text{total}$} \\ \hline
            12 & 18 & $0.1156$ & $\pm 0.0037$ & $\pm 0.0008$ & ${~}^{+0.0035}_{-0.0025}$ & $0.1525$ & $\pm 0.0086$ \\
            16 & 26 & $0.1153$ & $\pm 0.0026$ & $\pm 0.0006$ & ${~}^{+0.0028}_{-0.0017}$ & $0.1417$ & $\pm 0.0054$ \\
            19 & 35 & $0.1167$ & $\pm 0.0024$ & $\pm 0.0003$ & ${~}^{+0.0018}_{-0.0010}$ & $0.1363$ & $\pm 0.0039$ \\
            12 & 52 & $0.1164$ & $\pm 0.0032$ & $\pm 0.0002$ & ${~}^{+0.0011}_{-0.0003}$ & $0.1271$ & $\pm 0.0040$ \\
            11 & 84 & $0.1158$ & $\pm 0.0045$ & $\pm 0.0003$ & ${~}^{+0.0014}_{-0.0004}$ & $0.1172$ & $\pm 0.0047$ \\ \hline
            \end{tabular}
    \end{center}
    \caption{Values of the strong coupling determined using data at different scales $\mu$. Shown are the number of jet cross sections used in each determination, the representative scale $\langle\mu\rangle$ for each group and the value of $\alpha_s(M_Z^2)$ including all uncertainties from the fit. The last column shows the value of the strong coupling at the scale of the data $\alpha_s(\langle\mu\rangle^2)$ together with its combined and symmetrised uncertainty, as evolved using NNLO QCD.
    }
    \label{tab:running}
    \vfill
\end{table}

\begin{figure}[p]
    \vfill
    \begin{center}
        \includegraphics{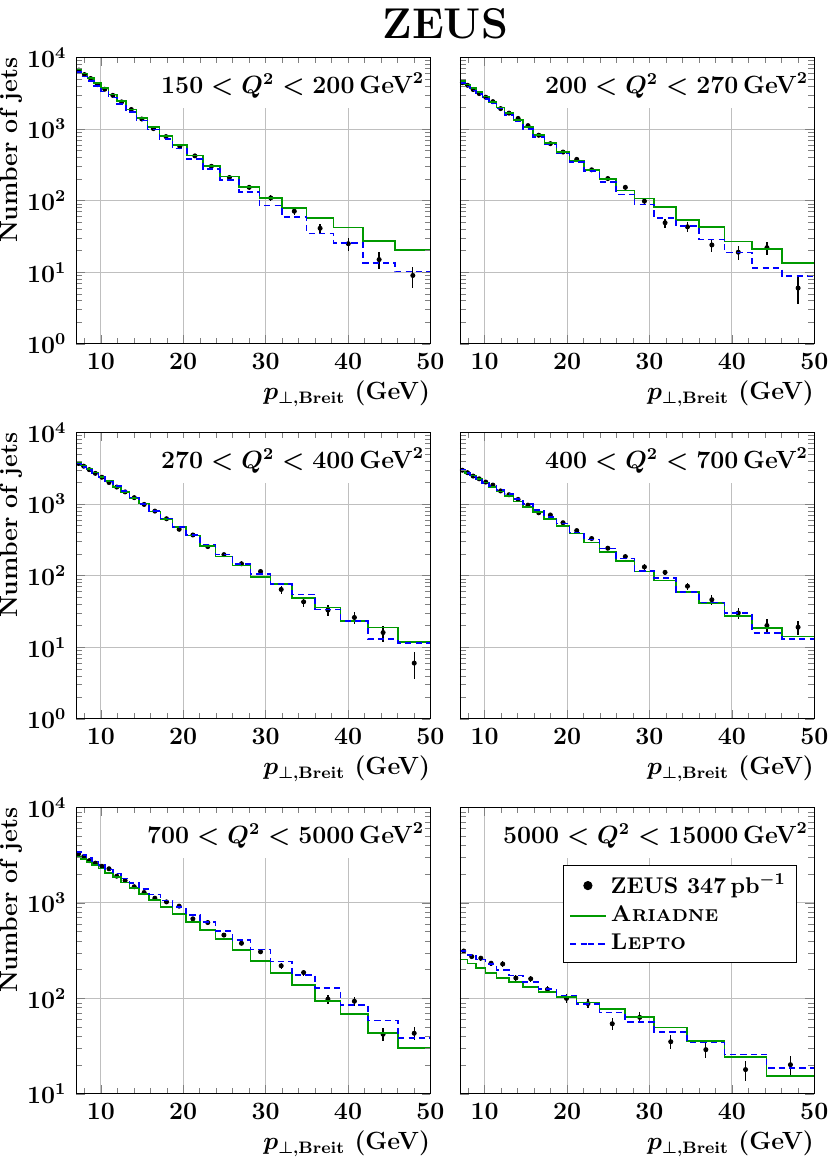}
    \end{center}
    \caption{Detector-level comparison of data (dots) and the \textsc{Ariadne} (solid, green) and \textsc{Lepto} (dashed, blue) MC distributions after corrections for the $p_{\perp,\text{Breit}}$ distribution in different regions of $Q^2$. The data are shown after subtracting the background from photoproduction and low-$Q^2$ DIS events. The error bars represent the statistical uncertainties of the data. The MC models are scaled globally to match the normalisation of the data in the fiducial range as defined in Sections~\ref{sec:selection} and \ref{sec:corrections}.}
    \label{fig:control jet}
    \vfill
\end{figure}

\begin{figure}[p]
    \vfill
    \begin{center}
        \includegraphics{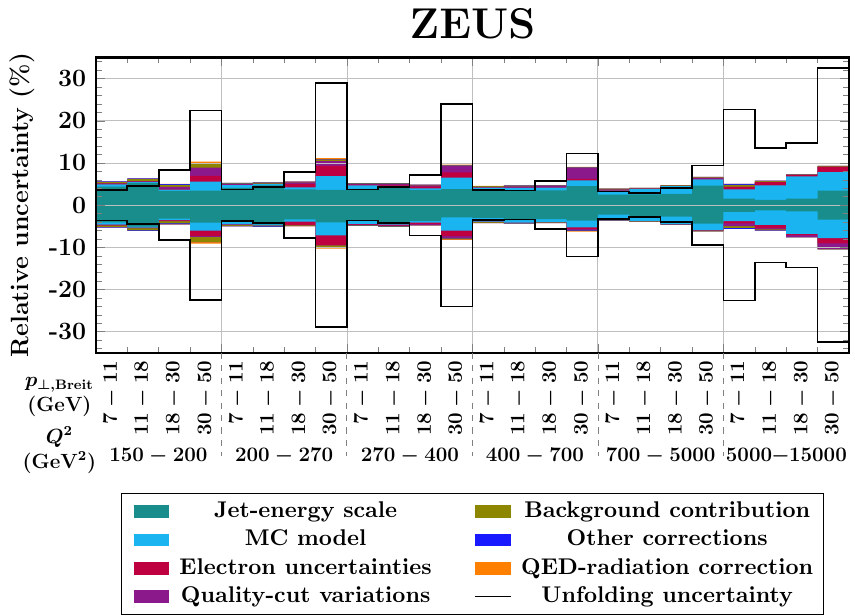}
    \end{center}
    \caption{Contributions of the different sources of systematic uncertainty, added in quadrature. The unfolding uncertainty is shown separately, without being added.
    The entry `MC model' includes the uncertainty due to exchanging the MC model ($\delta_\text{model}$) and the uncertainty in the reweighting of the MC models ($\delta_\text{rew.}$).
    The entry `Electron uncertainties' represents the sum of the uncertainties associated with the electron-energy scale ($\delta_\text{EES}$), electron-energy calibration ($\delta_\text{EL}$) and electron-finding algorithm ($\delta_\text{EM}$). Uncertainties due to photoproduction ($\delta_\text{PHP}$), low-$Q^2$ DIS ($\delta_\text{Low-$Q^2$}$) and unmatched jets ($\delta_\text{fake}$) are shown as the entry `Background contribution'.
    The polarisation uncertainty ($\delta_\text{pol.}$), track-association uncertainty ($\delta_\text{TME}$) and the uncertainty of the track reconstruction ($\delta_\text{FLT}$) are combined into the entry `Other corrections'.
    }
    \label{fig:systematics}
    \vfill
\end{figure}

\begin{figure}[p]
    \vfill
    \begin{center}
        \includegraphics{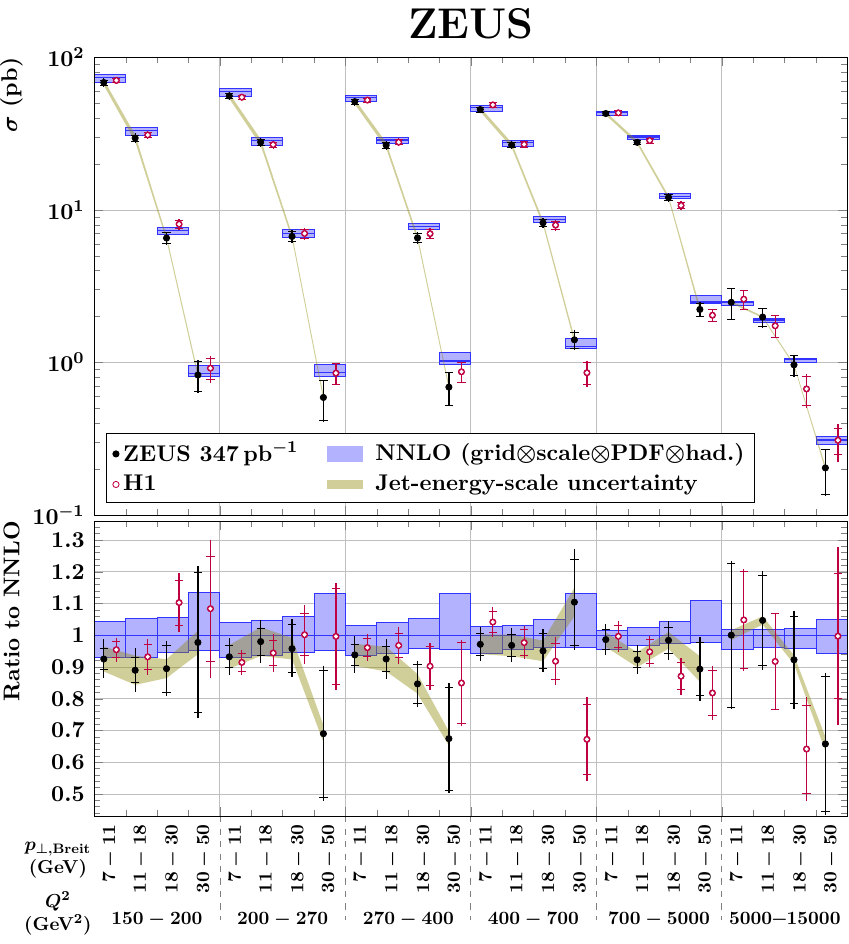}
    \end{center}
    \caption{The measured double-differential inclusive jet cross sections with $\unit[7]{\GeV} < p_{\perp,\text{Breit}} < \unit[50]{\GeV}$ and $-1 < \eta_\text{lab} < 2.5$, in the kinematic range $\unit[150]{\GeV^2} < Q^2 < \unit[15000]{\GeV^2}$ and $0.2 < y < 0.7$. Shown are the present measurement from ZEUS (full dots, black), the corresponding measurement from H1 (open dots, red) \protect\cite{h1highq2newjets} and the NNLO QCD predictions (blue boxes). The inner error bars of the measurements represent the unfolding uncertainty and the outer error bars the total uncertainty. For the ZEUS measurement, the shaded band shows the uncertainty associated with the jet-energy scale. The NNLO QCD calculation is computed at $\alpha_s(M_Z^2) = 0.1155$ using the HERAPDF2.0Jets NNLO PDF set and scales of $\mu_\text{r}^2 = \mu_\text{f}^2 = Q^2+p_{\perp,\text{Breit}}^2$. The predictions were corrected for hadronisation and for $Z$-boson exchange. Also shown is the ratio of those cross sections to the NNLO QCD predictions.
    }
    \label{fig:cross sections}
    \vfill
\end{figure}

\begin{figure}[p]
    \vfill
    \begin{center}
        \includegraphics{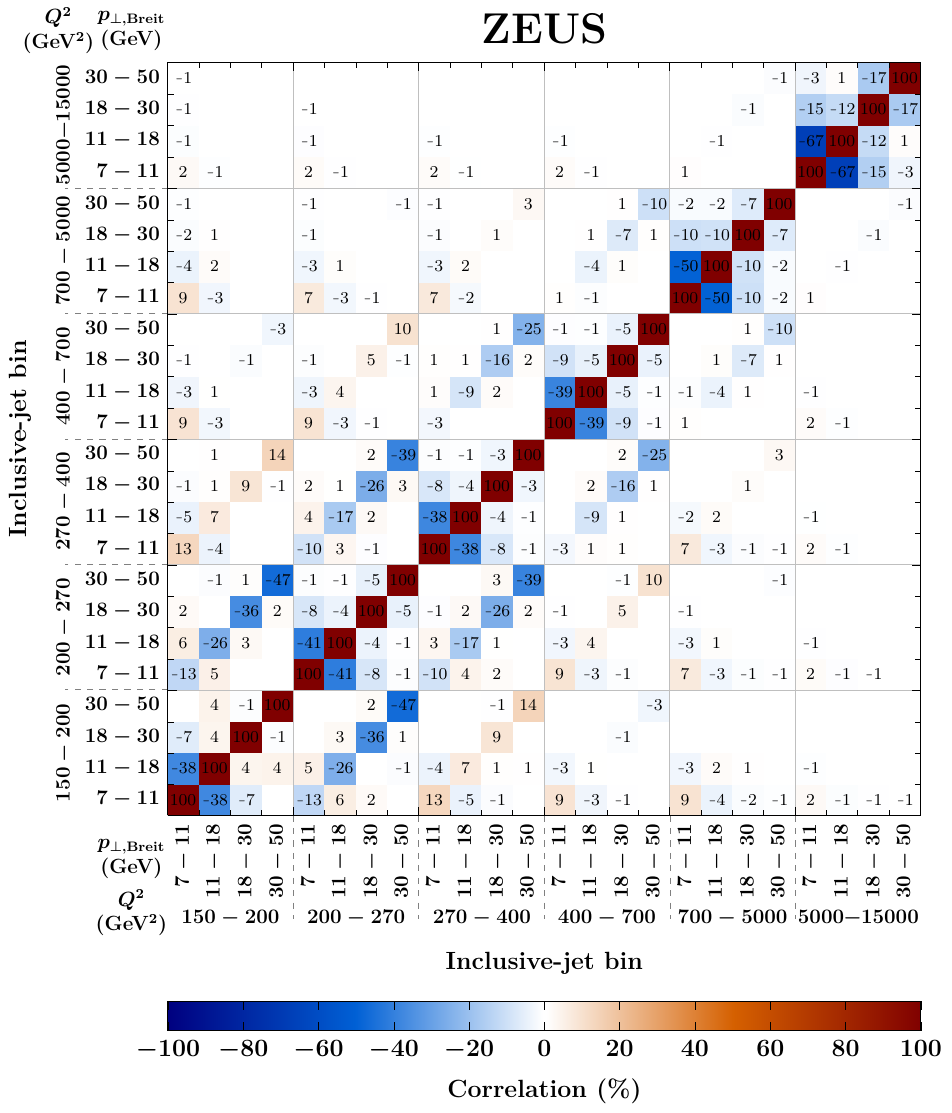}
    \end{center}
    \caption{Correlation matrix of the unfolding uncertainty for the inclusive-jet cross-section measurement. By definition, the matrix is symmetric and all entries on the diagonal are $100\%$. Negative correlations due to the finite detector resolution arise mostly in adjacent bins at small $Q^2$ and small $p_{\perp,\text{Breit}}$. Adjacent bins that do not belong to this region and non-adjacent bins are not strongly correlated.
    }
    \label{fig:correlation1}
    \vfill
\end{figure}

\begin{figure}[p]
    \vfill
    \begin{center}
        \includegraphics{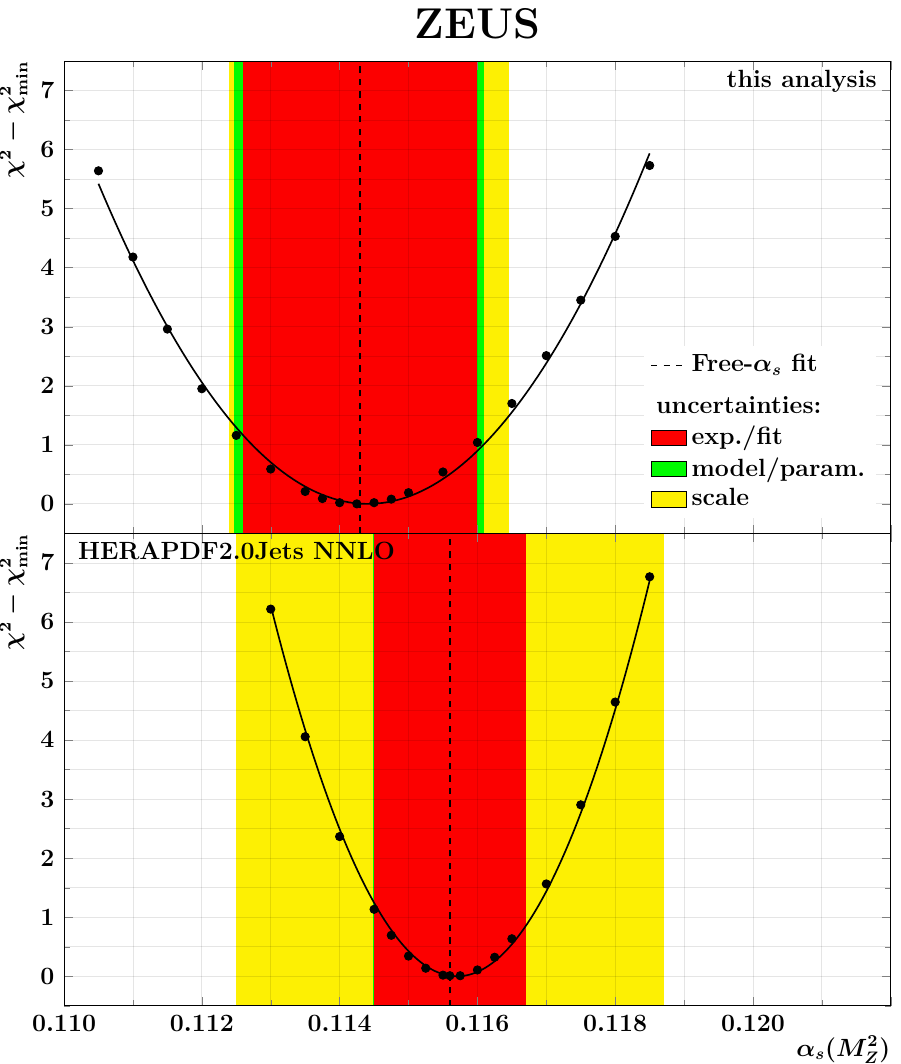}
    \end{center}
    \caption{Difference between $\chi^2$ and $\chi^2_\text{min}$ as a function of $\alpha_s(M_Z^2)$ for fits with fixed $\alpha_s(M_Z^2)$ at NNLO. The central value and the experimental/fit, model/parameterisation and scale uncertainties determined for the free $\alpha_s(M_Z^2)$-fit assuming fully correlated scale uncertainties are also shown, added in quadrature. For reference, the corresponding plot from the HERAPDF2.0Jets NNLO analysis is also shown\protect\cite{HERAPDF20NNLO}.}
    \label{fig:scan}
    \vfill
\end{figure}

\begin{figure}[p]
    \vfill
    \begin{center}
        \includegraphics{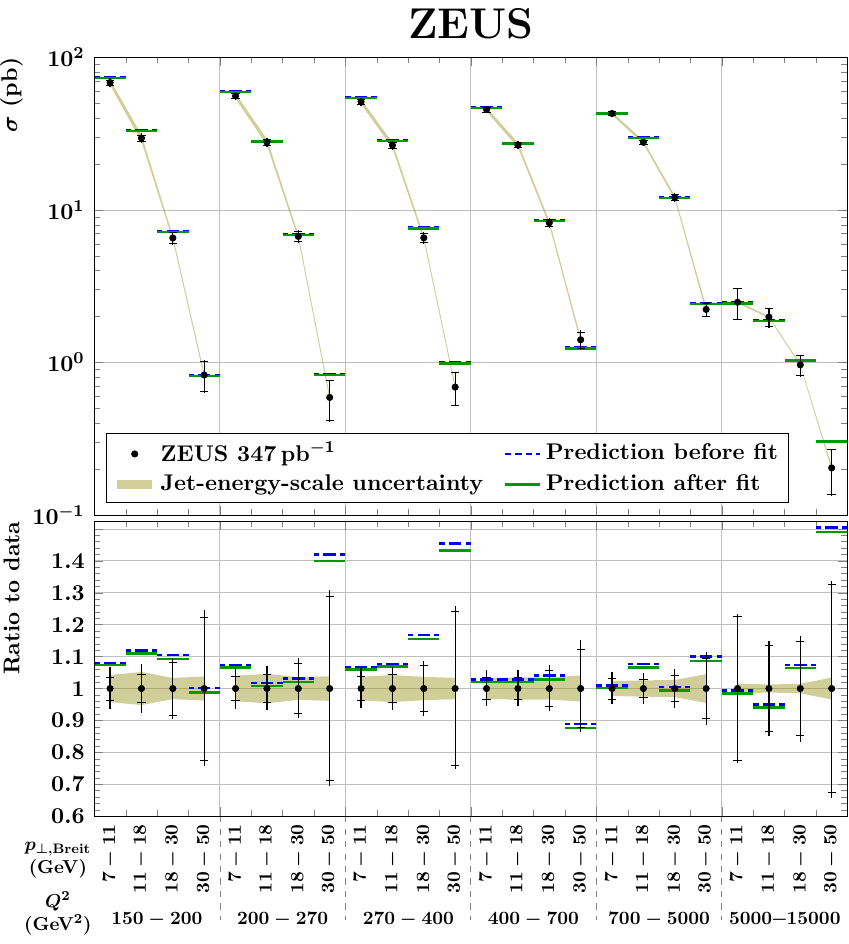}
    \end{center}
    \caption{Double-differential inclusive jet cross-section predictions based on the NNLO fit (solid, green) compared to the data (dots). Additionally, the predictions are shown before including the current inclusive-jet dataset in the fit (dashed, blue).
    The uncertainties of the fit results are not shown. When including the current dataset, the experimental/fit uncertainty decreases slightly.
    The ratios of the cross sections as calculated before and after the fit to the data are also shown. Other details as given in Fig.~\ref{fig:cross sections}.}
    \label{fig:fit nnlo}
    \vfill
\end{figure}

\end{document}